\documentclass[10pt,journal]{IEEEtran}              
\IEEEpeerreviewmaketitle
\ifCLASSOPTIONcompsoc
  \usepackage[nocompress]{cite}
\else
  \usepackage{cite}
\fi
\ifCLASSINFOpdf
  \usepackage[pdftex]{graphicx}
\else
  \usepackage[dvips]{graphicx}
\fi
\usepackage[cmex10]{amsmath}
\usepackage{amsmath,amsfonts,bm}
\usepackage{color,soul}
\usepackage[colorinlistoftodos,
           prependcaption,
           textsize=small,
           backgroundcolor=yellow,
           linecolor=lightgray,
           bordercolor=lightgray]{todonotes}

\usepackage{algorithm,algorithmic}

\usepackage{array}

\usepackage{mdwmath}
\usepackage{mdwtab}

\usepackage{eqparbox}

\ifCLASSOPTIONcompsoc
  \usepackage[caption=false,font=footnotesize,labelfont=sf,textfont=sf]{subfig}
\else
  \usepackage[caption=false,font=footnotesize]{subfig}
\fi
\usepackage{multirow}

\def\prodd{\displaystyle\prod}
\def\sumd{\displaystyle\sum}

\def\fracd{\displaystyle\frac}

\def\Sb{\mathbf{S}}
\def\Ab{\mathbf{A}}

\def\Xb{\mathbf{X}}
\def\Yb{\mathbf{Y}}

\def\Cb{\mathbf{C}}

\def\Pb{\mathbf{P}}
\def\xb{\mathbf{x}}
\def\xt{\mathbf{x}_{t}}
\def\yt{\mathbf{y}_{t}}

\def\nt{\mathbf{n}_{t}}

\def\Sigmab{\boldsymbol{\Sigma}}

\def\mub{\boldsymbol{\mu}}

\def\stm{s_{tm}}
\def\sttm{s_{(t-1)m}}
\def\xtm{x_{tm}}
\def\xttm{x_{(t-1)m}}

\def\s{\sigma}
\def\Ib{\mathbf{I}}
\def\1b{\mathbf{1}}

\def\slice{\vartheta}

\def\Acal{\mathcal{A}}
\def\Ccal{\mathcal{C}}

\def\Ncal{\mathcal{N}}
\def\Ocal{\mathcal{O}}
\def\Ucal{\mathcal{U}}
\def\Xcal{\mathcal{X}}

\newcommand{\xhdr}[1]{\vspace{1.7mm}\noindent{{\bf #1.}}}
\newcommand{\Htop}{\mathsf{H}}

\begin{document}
%
\title{Infinite Factorial Finite State Machine\\for Blind Multiuser Channel Estimation}
%
%
%

\author{Francisco~J.~R.~Ruiz,
        Isabel~Valera,
        Lennart Svensson,
        and~Fernando~Perez-Cruz
\IEEEcompsocitemizethanks{\IEEEcompsocthanksitem F.\ J.\ R.\ Ruiz is with the Department of Computer Science, Columbia University in the City of New York, United States of America, and with the Engineering Department, University of Cambridge, United Kingdom.\protect\\
E-mail: f.ruiz@columbia.edu
\IEEEcompsocthanksitem I.\ Valera is with the Department of Empirical Inference at the Max Planck Institute for Intelligent Systems in T\"{u}bingen, Germany, and with the Department of Signal Processing and Communications, University Carlos III in Madrid, Spain.
\IEEEcompsocthanksitem L.\ Svensson is with Chalmers University of Technology at Gothenburg, Sweden.
\IEEEcompsocthanksitem F.\ Perez-Cruz is Chief Data Scientist at the Swiss Data Science Center, Zurich, Switzerland, and is also with the Department of Signal Processing and Communications, University Carlos III in Madrid, Spain.
}
\thanks{Manuscript received MONTH DD, 2017; revised MONTH DD, 2017.}}

%
%

\markboth{IEEE TRANSACTIONS ON COGNITIVE COMMUNICATIONS AND NETWORKING}%
{Shell \MakeLowercase{\textit{et al.}}: Bare Advanced Demo of IEEEtran.cls for Journals}
%


\maketitle


\begin{abstract}
New communication standards need to deal with machine-to-machine communications, in which users may start or stop transmitting at any time in an asynchronous manner. Thus, the number of users is an unknown and time-varying parameter that needs to be accurately estimated in order to properly recover the symbols transmitted by all users in the system. In this paper, we address the problem of joint channel parameter and data estimation in a multiuser communication channel in which the number of transmitters is not known. For that purpose, we develop the infinite factorial finite state machine model, a Bayesian nonparametric model based on the Markov Indian buffet that allows for an unbounded number of transmitters with arbitrary channel length. We propose an inference algorithm that makes use of slice sampling and particle Gibbs with ancestor sampling. {Our approach is fully blind as it does not require a prior channel estimation step, prior knowledge of the number of transmitters, or any signaling information.} Our experimental results, loosely based on the LTE random access channel, show that the proposed approach can effectively recover the data-generating process for a wide range of scenarios{, with varying number of transmitters, number of receivers, constellation order, channel length, and signal-to-noise ratio}.
\end{abstract}


%
\IEEEpeerreviewmaketitle

\DeclareRobustCommand{\parhead}[1]{\noindent\textbf{#1.}~}

\ifCLASSOPTIONcompsoc
  \IEEEraisesectionheading{\section{Introduction}\label{sec:introduction}}
\else
  \section{Introduction}\label{sec:introduction}
\fi

\IEEEPARstart{O}{ne} of the trends in wireless communication networks (WCNs) is the increase of heterogeneity \cite{SmallCell2012}. {Nowadays,} users of WCNs are no longer only humans talking, and the number of services and users are booming. Machine-to-machine (M2M) communications and the Internet of Things (IoT) will shape the traffic in {WCN} in the years to come \cite{FocalPoint2003,Vodafone2010,Ericsson2011,Lien2011}. While there are millions of {M2M} cellular devices already using second, third and fourth generation cellular networks, the industry expectation is that the number of devices will increase ten-fold in the coming years \cite{Dhillon2014}.

{M2M} traffic, which also includes communication between a sensor/actuator and a corresponding application server in the network, is distinct from consumer traffic, which has been the main driver for the design of long term evolution (LTE) systems. First, while current consumer traffic is characterized by small number of long lived sessions, {M2M} traffic involves a large number of short-lived sessions, with typical transactions of a few hundred bytes. The short payloads involved in {M2M} communications make it highly inefficient to establish dedicated bearers for data transmission. Therefore, in some cases it is better to transmit small payloads in the random access request itself \cite{Chen2010}. Second, a significant number of battery powered devices are expected to be deployed at adverse locations such as basements and tunnels (e.g., underground water monitors and traffic sensors) that demand superior link budgets. Motivated by this need for increasing the link budget for {M2M} devices, transmission techniques that minimize the transmit power for short burst communications are needed \cite{Dhillon2013}. Third, the increasing number of {M2M} devices requires new techniques on massive access management \cite{Tu2011,Ho2012}. Due to these differences, there are strong arguments for the need to optimize WCNs specifically for {M2M} communications \cite{Dhillon2014}. 

The nature of {M2M} traffic leads to multiuser communication systems in which a large number of users may aim at entering or leaving the system (i.e., start or stop transmitting) at any given time. In this context, we need a method that allows the users to access the system in a way that the signaling overhead is reduced. {The method should determine the number of users transmitting in a communication system, and jointly perform channel estimation and detect the transmitted data, with minimum pilot signaling.}
This problem appears in several specific applications. For instance, in the context of wireless sensor networks, where the communication nodes can often switch on and off asynchronously during operation. It also appears in massive multiple-input multiple-output (MIMO) multiuser communication systems \cite{Hoydis2013,Lu2014}, in which the base station has a very large number of antennas and the mobile devices use a single antenna to communicate within the network. In a code division multiple access (CDMA) context, a set of terminals randomly access the channel to communicate with a common access point, which receives the superposition of signals from the active terminals only \cite{Vazquez2013}. In these applications, the number of users is an unknown and time-varying parameter that we need to infer.

In this paper, we aim at solving the channel estimation and symbol detection problems in a fully unsupervised way, without the need of signaling data. {Our approach is thus suitable} for applicability on the random access channel, in which more than one terminal may decide to transmit data at the same time. To this end, we advocate for the use of Bayesian nonparametric (BNP) tools, which can adapt to heterogeneous structures by considering an unbounded number of users in their prior. We develop a BNP model that has the required flexibility to account for any number of transmitters without the need of additional previous knowledge or bounds. We are interested in showing that BNP models can solve the problem blindly and that versatility can improve detection in heterogeneous WCNs.
Thus, or goal is to show that we can recover the messages and infer the number of transmitters using as little information as possible. The transmitted messages are therefore assumed to lack additional pilots or other structure. This makes our approach applicable on any interference-limited multiuser communication system in which the transmission of pilots or the need for synchronization may severely limit the spectral efficiency of the WCN.

The main difference between our model and the existing approaches in the literature is that our model is fully blind{---with respect to the channel coefficients, the number of transmitters, and the transmitted symbols---}and can recover the number of users without the need of synchronization. Indeed, the problem of multi-user detection has been widely studied in the literature, using standard linear methods (least squares or minimum mean squared error) or non-linear approaches (e.g., successive interference cancellation or parallel interference cancellation) \cite{Verdu1998,Jiang2007}. Most of these approaches require synchronization techniques (see, e.g., \cite{Wu1998,Pad2011,Vazquez2013}). In order to perform the user identification step, the common approach is to pre-specify signature sequences that are specific for each user \cite{Pi2002,Buzzi2010,Mousavi2011}, which imposes a constraint on the total number of users in the system.

{Our model also differs from existing approaches in other ways. For example, in} \cite{Halford1998}{ users may only enter or leave the system at certain pre-defined instants. In} \cite{Angelosante2009b}{ the user identification step is performed first without symbol detection. The approach in} \cite{Zhu2011}{ requires knowledge of the channel, and the method in} \cite{Vazquez2013} {is restricted to flat-fading channels. The methods in} \cite{Angelosante2009a,Angelosante2010}{ come at the cost of both important approximations and prohibitive computational cost (experiments are presented with short frames of only ten symbols). In contrast, our model is fully blind, it does not need a previous channel estimation step, and it is not restricted to flat-fading channels or very short frames.}

All these methods {assume that the number of users in the system is known,} which makes sense in a {DS-CDMA} system but may represent a limitation in other scenarios. {They also require synchronization mechanisms. We build on previous BNP approaches that do not have these constraints} \cite{Valera2014,Valera2015,Valera2015eusipco}{. In particular, we extend the method in} \cite{Valera2015eusipco}{ to channels with memory, and we address the exponential complexity of} \cite{Valera2014,Valera2015}{, which is only applicable for BPSK systems and for a channel length of at most $2$. {Instead, the complexity of our algorithm scales with the square of the channel length.}}

{In summary, our model does not need any of the following requirements: a constraint on the number of users, that the users are synchronized with the network, that the users transmit a preamble, that the channel is known, or that the channel length equals one.} Our model allows for an unbounded number of transmitters due to its nonparametric nature, and it is not restricted to memoryless channels because transmitters are modeled as finite state machines. Our goal is to show that we can recover the number of transmitters and their payloads with very little information.

Our model {follows the three steps of cognitive communications. It gathers the incoming signal, which is a mixture of all transmitters that are active at a given time instant (perception); it learns the number of users, the channel they face, and it detects their payload in a blind way (learning); and it assigns the symbols from each user and passes that information to the network to act upon (decision-making). These features} can mitigate the effect of collisions and reduce the number of retransmissions over the random access channel, which is of mayor concern in LTE-A systems when M2M communications start being a driving force in WCNs \cite{Ali16, Polese16}.

\subsection{Technical contributions}

We model the multiuser communication system as an infinite factorial finite state machine (IFFSM) in which a potentially infinite number of finite state machines (FSMs), each representing a single transmitter, contribute to generate the observations. The symbols transmitted by each user correspond to the inputs of each FSM, and its memory accounts for the multipath propagation between each transmitter-receiver pair. The output of the IFFSM corresponds to the received signal, which depends on the inputs and the current states of the active FSMs (i.e., the active transmitters), under noise. 
Our IFFSM considers that transmitters may start or stop transmitting at any time, and it ensures that only a finite subset of the users become active during any finite observation period, while the remaining (infinite) transmitters remain in an idle state (i.e., they do not transmit). 

As for most BNP models, one of the main challenges of our IFFSM is posterior inference. We develop a suitable inference algorithm by building a Markov chain Monte Carlo (MCMC) kernel using particle Gibbs with ancestor sampling (PGAS) \cite{Lindsten2014}, a recently proposed algorithm that belongs to the broader family of particle Markov chain Monte Carlo (PMCMC) methods \cite{Andrieu2010}. 
This algorithm presents quadratic complexity with respect to the memory length, avoiding the exponential complexity of previous approaches, such as forward-filtering backward-sampling schemes \cite{Ghahramani1997,VanGael2009,Valera2014,Valera2015}.
Our experimental results, based on the LTE random access channel, show that the proposed approach efficiently solves user activity detection, channel estimation and data detection in a jointly and fully blind way.


\subsection{Organization of the paper}

The rest of the paper is organized as follows. In Section~\ref{sec:Background}, {we review the basic building blocks} that we use to develop our model, namely, the stochastic finite-memory FSM and the Markov Indian buffet process (mIBP). Section~\ref{sec:model} details our proposed IFFSM, whereas we describe the inference algorithm in Section~\ref{sec:inference}. Sections~\ref{sec:experiments} and \ref{sec:conclusions} are devoted to the experiments and conclusions, respectively.


\section{Background}\label{sec:Background}


{Here we describe the two main building blocks that we use to develop our IFFSM:} the FSM \cite{Hopcroft2006} and the mIBP \cite{VanGael2009}.

\subsection{Stochastic Finite State Machines}\label{sec:FSM}

FSMs have been applied to a huge variety of problems in many areas, including biology (e.g., neurological systems), signal processing (e.g., speech modeling), control, communications and electronics \cite{Anderson2006,Wang2012}. In its general form \cite{Hopcroft2006}, an {FSM} is defined as an abstract machine consisting of:
\begin{itemize}
\item A set of states, which might include one or several initial and final states.
\item A set of discrete or continuous inputs.
\item A set of discrete or continuous outputs.
\item A state transition function, which takes the current input and state and returns the next state.
\item An emission function, which takes the current state and input and returns an output.
\end{itemize}

\begin{figure}[t]
\centering
\subfloat[]{\includegraphics[width=4.9cm]{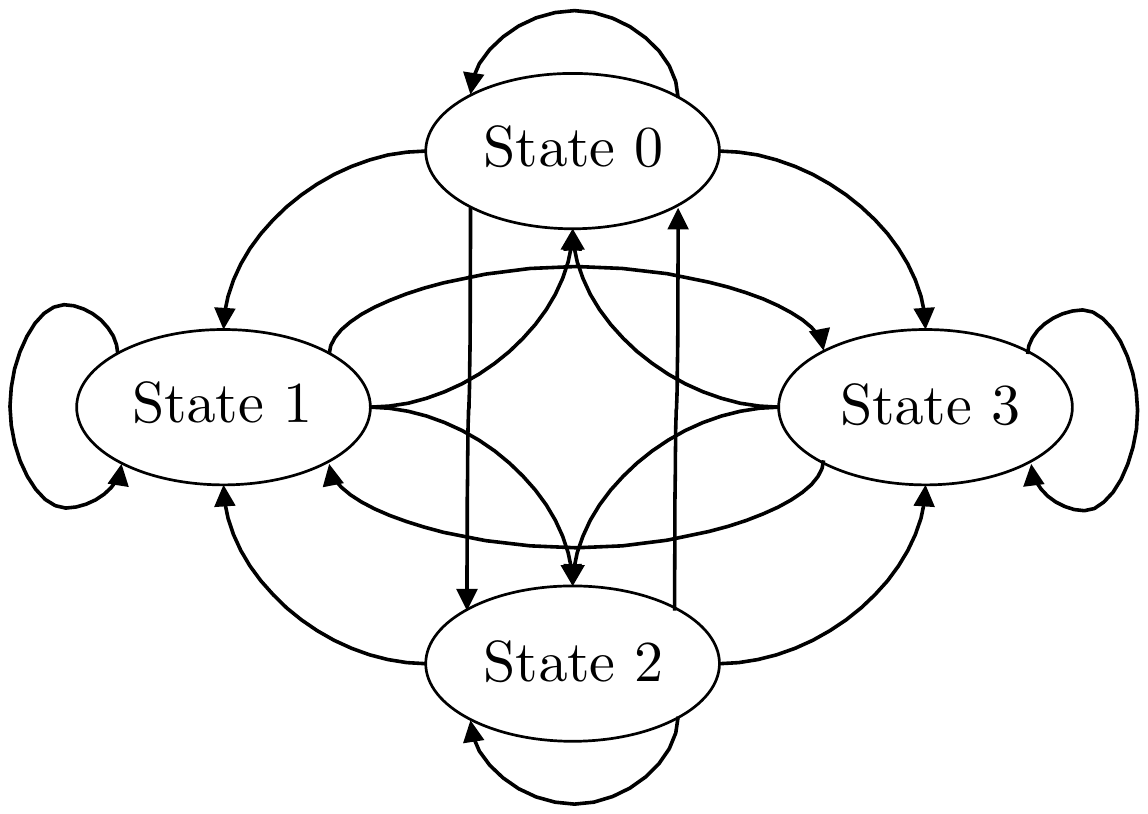}}
\subfloat[]{\includegraphics[width=4.2cm]{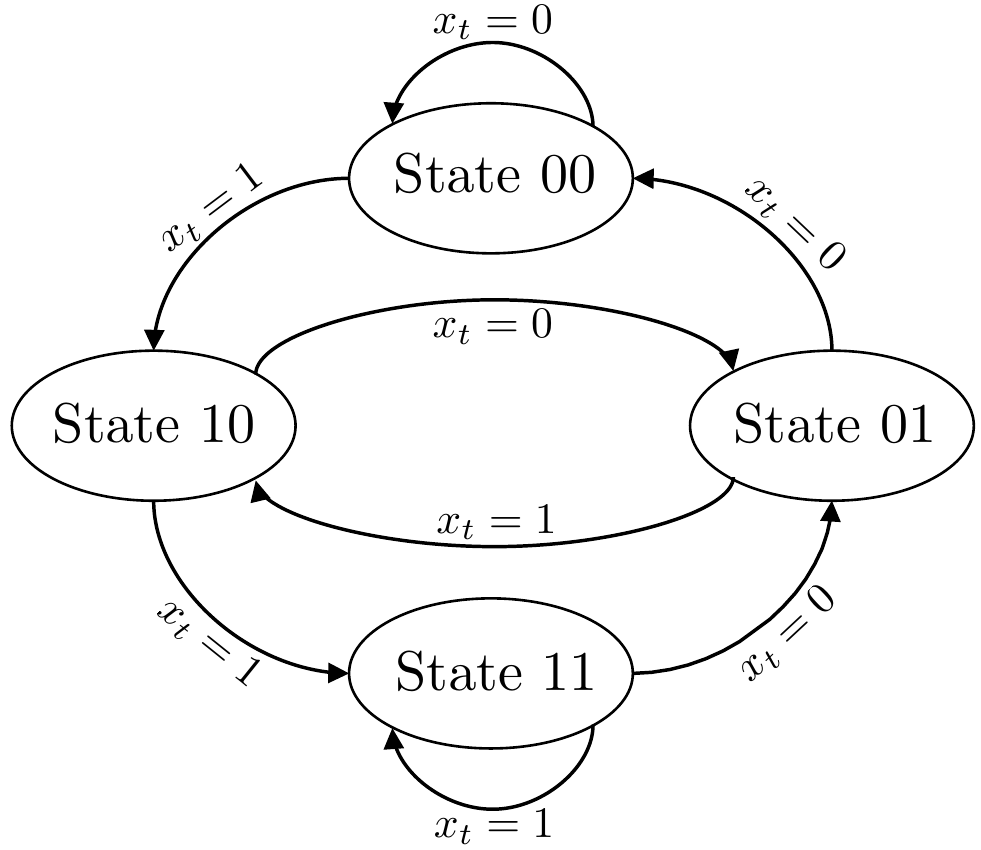}\label{fig:ballsFSM}}
\caption[State diagram of an HMM and an FSM]{(a) State diagram of an HMM with $4$ states, in which all transitions among states are allowed. (b) State diagram of a stochastic finite-memory FSM with memory length $L=2$ and $\Xcal=\{0,1\}$. Each state can be represented by the vector containing the last $L$ input events. Each state can only transition to other $|\Xcal|=2$ states, depending on the input event $x_t$. Hence, not all transitions among states are allowed, but it is possible to reach any other state in the graph after $L=2$ transitions.}
\label{fig:balls}
\end{figure}

The transition and emission functions are, in general, stochastic (e.g., hidden Markov models (HMMs) \cite{Baum1966,Rabiner1986}), which implies that the input events and the states are not directly observable through the output events (instead, each input and state produces one of the possible output events with a certain probability). We focus on stochastic finite-memory FSMs, in which each state can be represented as a finite-length sequence containing the last $L$ inputs. This {FSM} relies on a finite memory $L$ and a finite alphabet $\Xcal$. Each input $x_t\in\Xcal$ produces a deterministic change in the state of the {FSM}, and a stochastic observable output $\yt$ at time $t$. The next state and the output depend on the current state and the input. The state can be expressed as the vector containing the last $L$ inputs, i.e., $\left[ x_t,x_{t-1},\ldots,x_{t-L+1}\right]$, therefore yielding $|\Xcal|^L$ different states, where $|\Xcal|$ denotes the cardinality of the set $\Xcal$. Each state can only transition to $|\Xcal|$ different states, depending on the next input event. A state diagram of an HMM and an FSM is depicted in Figure~\ref{fig:balls}.

\begin{figure}[t]
\includegraphics[width=5cm]{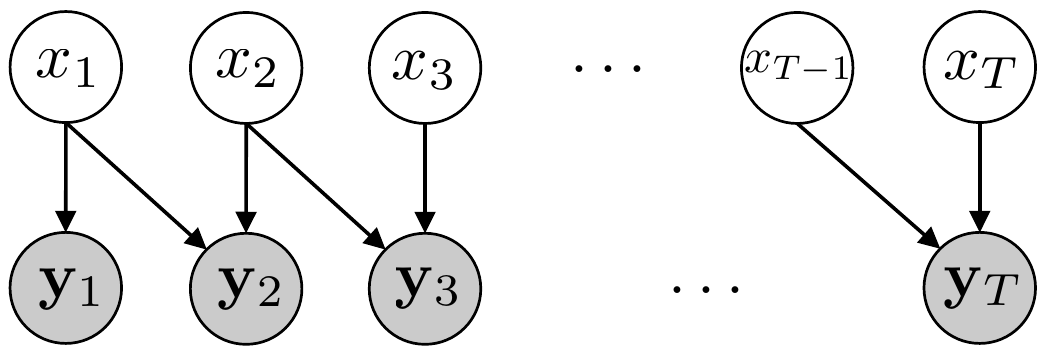}
\centering
\caption{Graphical representation of an FSM model with memory length $L=2$. Here, $x_t$ denotes the transmitted symbol at time $t$, and $\mathbf{y}_t$ is the received (observed) signal. The arrows indicate that the $t$-th received symbol depends on both the $t$-th and the ($t-1$)-th transmitted symbols.}
\label{ch1:fig:fsm}
\end{figure}

For any finite-memory FSM, the probability distribution over the inputs $x_t$ and the observations (outputs) $\yt$ can be written as
\begin{equation}
p(\{x_t,\yt\}_{t=1}^T) = \prodd_{t=1}^T p(x_{t}) \prodd_{t=1}^T p(\yt|x_{t},\ldots,x_{t-L+1}),
\end{equation}
i.e., the likelihood of each observation $\yt$ depends not only on $x_t$, but also on the previous $L-1$ inputs. The model also requires the specification of the initial state, which is defined by the inputs $x_0,\ldots,x_{1-L}$. We show the graphical model for an {FSM} with memory length $L=2$ in Figure~\ref{ch1:fig:fsm}. Note that this model can be equivalently represented as a standard HMM with a sparse transition probability matrix of size $|\Xcal|^L\times |\Xcal|^L$. However, the HMM representation of the FSM leads to a computationally intractable inference algorithm due to the exponential dependency on the memory length $L$ of the state space cardinality.


\subsection{Markov Indian Buffet Process}\label{subsec:mIBP}
The central idea behind BNPs is the replacement of classical finite-dimensional prior distributions with general stochastic processes, allowing for an open-ended number of degrees of freedom in a model \cite{Jordan2010}. They constitute an approach to model selection and adaptation in which the model complexity is allowed to grow with data size \cite{Orbanz2010}. 

Within the family of BNP models, the Markov Indian buffet process (mIBP) is a variant of the Indian buffet process (IBP) \cite{Griffiths2011} and constitutes the main building block of the infinite factorial hidden Markov model (IFHMM), which considers an infinite number of first-order Markov chains with binary-valued states that evolve independently \cite{VanGael2009}.

The mIBP places a prior distribution over a binary matrix $\Sb$ with a finite number of rows and an infinite number of columns. The $t$-th row represents time step $t$, whereas the $m$-th column contains the (binary) states of the $m$-th Markov chain. Each element $\stm=(\Sb)_{tm}\in\{0,1\}$ indicates whether the $m$-th Markov chain is active at time instant $t$, with $t=1,\ldots,T$, $m=1,\ldots,M$, and $M\rightarrow\infty$. The states $\stm$ evolve according to the transition matrix
\begin{equation}
\Ab^m=\left(
			 \begin{array}{cc}
			 1-a^m & a^m \\
			 1-b^m & b^m
			 \end{array}
		     \right),
\end{equation}
i.e., $a^m=p(s_{tm}=1|s_{(t-1)m}=0)$ is the transition probability from inactive to active, and $b^m=p(s_{tm}=1|s_{(t-1)m}=1)$ is the self-transition probability of the active state. We assume an inactive initial state for all chains at time $t = 0$.

In order to specify the prior distribution over the transition probabilities $a^m$ and $b^m$, we focus here on the stick-breaking construction of the IBP in \cite{Teh2007}, because it allows for simple and efficient inference algorithms. Under this construction, we first introduce the notation $a^{(m)}$ to denote the sorted values of $a^m$, such that $a^{(1)}>a^{(2)}>a^{(3)}>\ldots$, {with}
\begin{equation}\label{eq:stick_a1}
a^{(1)}\sim \textrm{Beta}(\alpha,1),
\end{equation}
and
\begin{equation}
p(a^{(m)}| a^{(m-1)}) \propto (a^{(m)})^{\alpha-1}\mathbb{I}(0\leq a^{(m)} \leq a^{(m-1)}),
\end{equation}
being $\mathbb{I}(\cdot)$ the indicator function, which takes value one if its argument is true and zero otherwise. Here, $\alpha$ is the concentration hyperparameter of the model, which controls the expected number of latent Markov chains that become active \emph{a priori}.
The prior over variables $b^m$ is given by
\begin{equation}\label{eq:stick_bm}
b^m\sim \textrm{Beta}(\beta_0,\beta_1),
\end{equation}
which does not depend on the index $m$
. In \eqref{eq:stick_a1} and \eqref{eq:stick_bm}, $\alpha$, $\beta_0$ and $\beta_1$ are hyperparameters of the model. 
This prior distribution ensures that, for any finite value of $T$, only a finite number of columns $M_+$ become active, while the rest of them remain in the all-zero state.


\section{Model Description}\label{sec:model}

We use the stochastic finite-memory FSM and the mIBP detailed in Section~\ref{sec:Background} as building blocks for our IFFSM model, which we use for modeling a multiuser communication system in which the number of transmitters is potentially infinite.

The key idea of our approach is to model the multiuser communication system as a factorial FSM in which each FSM represents a user, and the inputs to the FSM are the symbols sent by each transmitter. The memory of the FSM accounts for the multipath propagation between each transmitter-receiver pair, and the output of the factorial FSM model corresponds to the received signal, which depends on the inputs and the current states of all the active FSMs (transmitters).

In order to account for an unbounded number of transmitters (parallel FSMs), we need to consider an inactive state, such that the observations cannot depend on those transmitters that are inactive. Furthermore, we need to ensure that only a finite number of them become active for any finite-length observed sequence. As detailed in Section~\ref{subsec:mIBP}, the mIBP in \cite{VanGael2009} meets these requirements by placing a prior distribution over binary matrices with an infinite number of columns. 

\subsection{Infinite Factorial Finite State Machine}\label{subsec:IFFSM}

We place a mIBP prior over an auxiliary binary matrix $\Sb$, where each element $\stm$ indicates whether the $m$-th FSM is active at time instant $t$. While active, the input symbol to the $m$-th {FSM} at time instant $t$, denoted by $\xtm$, is assumed to belong to the set $\Acal$, with finite cardinality $|\Acal|$. Here, $\Acal$ represents the constellation, and $\xtm$ stands for the transmitted symbol of the $m$-th transmitter at time instant $t$. We assume that the transmitted symbols are independent and uniformly distributed on the set $\Acal$. While inactive, we can assume that $\xtm = 0$ and, therefore, each input $\xtm \in \Xcal$, with $\Xcal=\Acal\bigcup\{0\}$. This can be expressed as
\begin{equation}\label{ch4:eq:p_xtm}
\xtm|\stm \sim \left\{ \begin{array}{ll}
                           \delta_0(\xtm) & \textrm{if } \stm=0,\\
                           \Ucal(\Acal) & \textrm{if } \stm=1,\\
                           \end{array}\right.
\end{equation}
where $\delta_0(\cdot)$ denotes a point mass located at $0$, and $\Ucal(\Acal)$ denotes the uniform distribution over the set $\Acal$.

In the IFFSM model, each input symbol $\xtm$ does not only influence the observation $\yt$ at time instant $t$, but also the future $L-1$ observations, $\mathbf{y}_{t+1},\ldots,\mathbf{y}_{t+L-1}$. Therefore, the likelihood function for $\yt$ depends on the last $L$ input symbols of all the FSMs, yielding
\begin{equation}\label{ch4:eq:likelihood}
p(\yt|\Xb) = p(\yt|\{ x_{tm},x_{(t-1)m},\ldots,x_{(t-L+1)m}\}_{m=1}^{\infty}),
\end{equation}
with $\Xb$ being the $T\times M$ matrix that contains all symbols $\xtm$. We assume innactive states $\stm=0$ for $t\leq 0$ (note that $\stm=0$ implies $\xtm=0$ and viceversa). 

\begin{figure}[t]
\includegraphics[width=7cm]{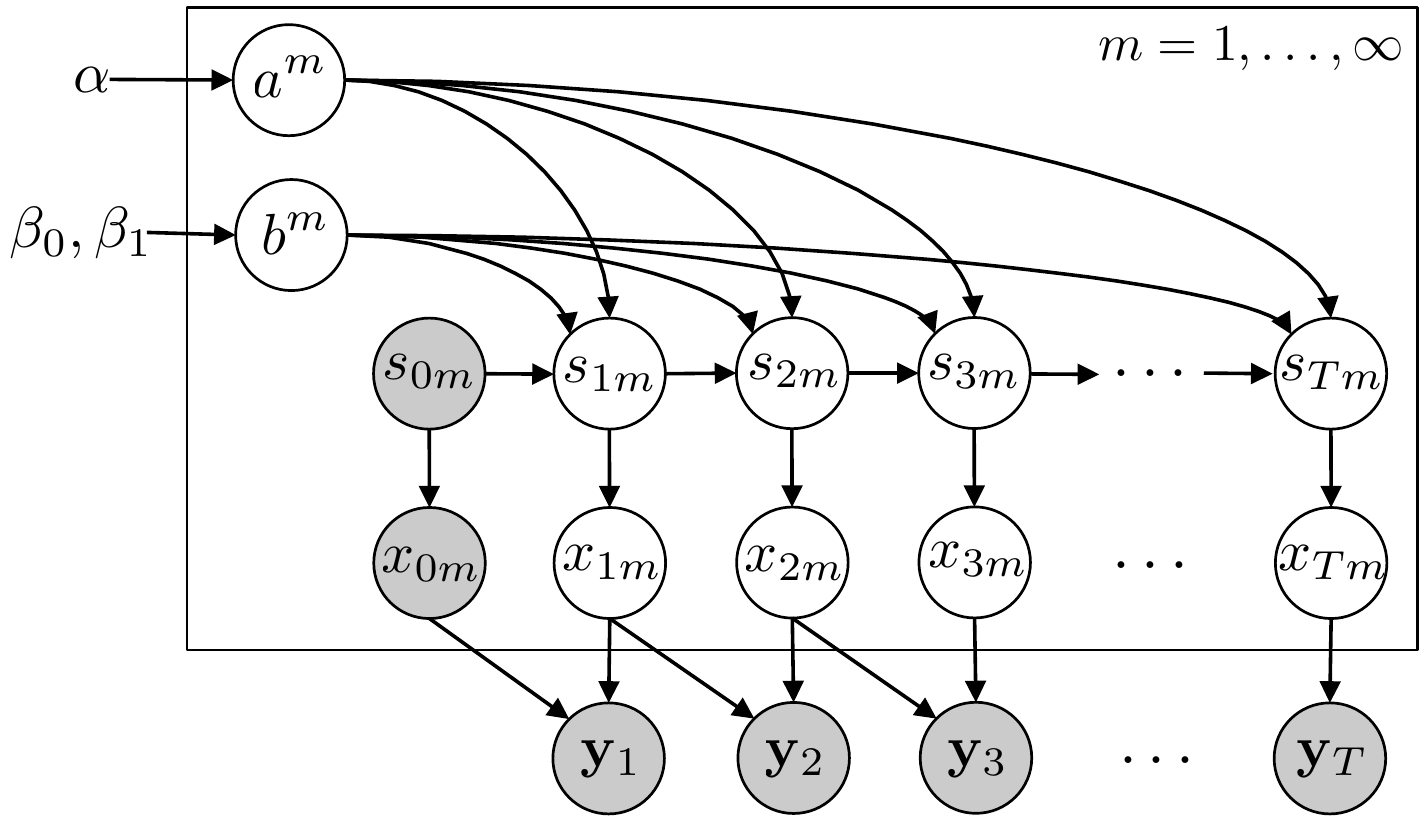}
\centering
\caption{Graphical model of the IFFSM with memory length $L=2$. Here, $x_{tm}$ denotes the transmitted symbol by device $m$ at time instant $t$, and $\mathbf{y}_t$ are the observations (a combination of the signals transmitted by all the active devices). The variable $s_{tm}$ indicates whether device $m$ is active or inactive at time instant $t$. The variables $a^m$ and $b^m$ govern the transition probabilities from inactive to active and vice-versa. The hyperparameters of the model are $\alpha$, $\beta_0$, and $\beta_1$.}
\label{ch4:fig:graphicalModel}
\end{figure}

The resulting {IFFSM} model, particularized for $L=2$, is shown in Figure~\ref{ch4:fig:graphicalModel}. Note that this model can be equivalently represented as a non-binary version of the IFHMM in \cite{VanGael2009}, using the extended states given by the vector $[x_{tm}, x_{(t-1)m}, \ldots, x_{(t-L+1)m}].$
However, we maintain the representation in Figure~\ref{ch4:fig:graphicalModel} because it allows deriving an efficient inference algorithm.

\subsection{Observation Model}\label{subsec:obsModel}

The model described in the previous section is general and can be applied to any sequence that can be explained by a potentially unbounded number of parallel FSMs. In this section, we particularize this model for its applicability on wireless communications, in which the parallel chains correspond to different transmitters. The {IFFSM} requires two general conditions for the likelihood model to be valid as the number of FSMs $M$ tends to infinity: i) the likelihood must be invariant to permutations of the chains, and ii) the distribution on $\yt$ cannot depend on any parameter of the $m$-th {FSM} if $s_{\tau m} = 0$ for $\tau=t-L+1,\ldots,t$. These conditions are met by simultaneous transmissions in WCNs. The first one says that the likelihood does not depend on how we label the different transmitters, and the second one indicates that inactive transmitters should not affect the observations. Among others, the standard discrete-time interference channel model with additive white Gaussian noise fulfils these restrictions:
\begin{equation}\label{ch4:eq:yt}
\yt = \sumd_{m=1}^{M_+} \sumd_{\ell=1}^{L} \mathbf{h}_m^\ell x_{(t-\ell+1)m}+\nt.
\end{equation}
In \eqref{ch4:eq:yt}, $x_{tm}$ represents the complex input (i.e., a symbol from a given constellation) at time instant $t$ for the $m$-th {FSM} (user), $\mathbf{h}_m^\ell$ are the channel coefficients (emission parameters, in the BNP literature), and $\nt$ is the additive white noise, which is circularly symmetric complex Gaussian distributed,\footnote{The complex Gaussian distribution $\Ccal\Ncal(\mub,\bm{\Gamma},\mathbf{C})$ over a vector $\xb$ of length $D$ is given by
$p(\xb)=\frac{1}{\pi^{D}\sqrt{\det(\bm{\Gamma})\det(\Pb)}}$ $\times 
 \exp\left\{ -\frac{1}{2} \left[(\xb-\mub)^\Htop,\hspace*{2pt}(\xb-\mub)^{\top}\right]
\left[ \begin{array}{cc} \bm{\Gamma} & \Cb \\ \Cb^\Htop & \bm{\Gamma}^{\star}\end{array}\right]^{-1}
\left[ \begin{array}{c} \xb-\mub \\ (\xb-\mub)^{\star} \end{array}\right] \right\}$, 
where $\Pb=\bm{\Gamma}^{\star}-\Cb^\Htop \bm{\Gamma}^{-1} \Cb$, 
$(\cdot)^\star$ denotes the complex conjugate, and $(\cdot)^{\Htop}$ denotes the conjugate transpose. A circularly symmetric complex Gaussian distribution has $\mub=\bm{0}$ and $\Cb=\bm{0}$.} i.e.,
\begin{equation}\label{ch4:eq:p_nt}
\nt \sim\Ccal\Ncal(\mathbf{0},\s^2_y\Ib_D,\mathbf{0}).
\end{equation}
The hyperparameter $\s^2_y$ is the noise variance, $\Ib_D$ is the identity matrix of size $D$, and $D$ is the number of receiving antennas, and hence the dimensionality of $\yt$, $\mathbf{h}_m^\ell$ and $\nt$. Consequently, the probability distribution over the received complex vector at time $t$ is described by
\begin{equation}
\yt|\{\mathbf{h}_m^\ell\},\Xb \sim \Ccal\Ncal\left(\sumd_{m=1}^{M_+} \sumd_{\ell=1}^{L} \mathbf{h}_m^\ell x_{(t-\ell+1)m},\s^2_y\Ib_D,\mathbf{0}\right).
\end{equation}

\begin{figure}[t]
\includegraphics[width=7.7cm]{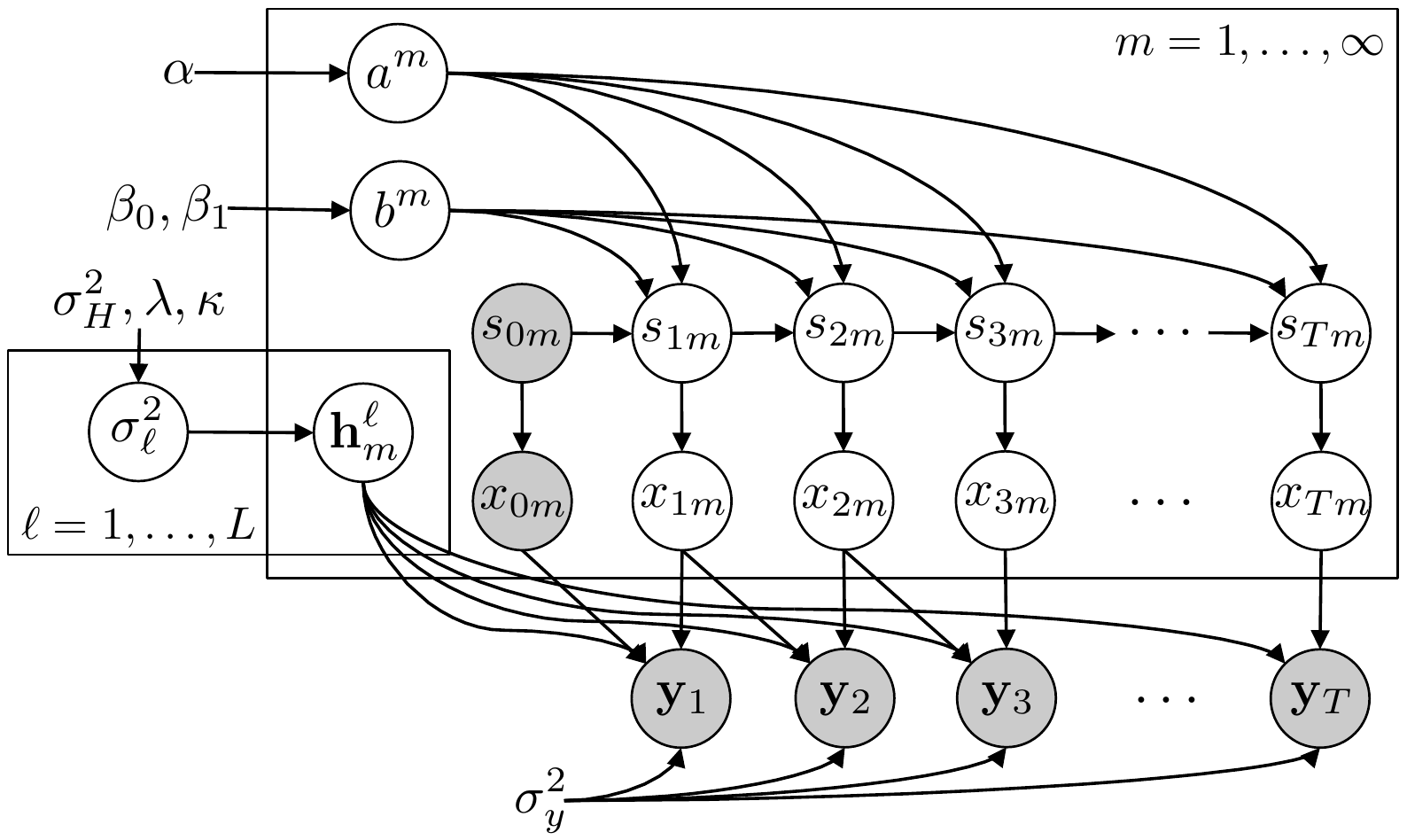}
\centering
\caption{Graphical Gaussian observation model for an IFFSM with memory length $L=2$. It combines the building block from Figure \ref{ch4:fig:graphicalModel} with the channel model coefficients $\mathbf{h}_m^{\ell}$, the prior variance of those channel coefficients $\sigma_\ell$, and the noise variance $\sigma_y^2$.}
\label{ch4:fig:graphicalModelMIMO}
\end{figure}

We finally place a circularly symmetric complex Gaussian prior over the channel coefficients, i.e.,
\begin{equation}\label{ch4:eq:p_hm}
\mathbf{h}_m^\ell | \sigma_{\ell}^2 \sim\Ccal\Ncal(\mathbf{0},\sigma_{\ell}^2\Ib_D,\mathbf{0}),
\end{equation}
and an inverse-gamma hyper-prior over the variances $\sigma_{\ell}^2$,
\begin{equation}
p(\s^2_{\ell})=
\frac{(\nu_\ell)^{\tau}}{\Gamma (\tau)} \left(\frac{1}{\s^2_{\ell}}\right)^{\tau+1} \exp\left\{-\frac{\nu_\ell}{\s^2_{\ell}}\right\},
\end{equation}
where $\nu_\ell=(\tau-1)\s^2_H e^{-\lambda(\ell-1)}$ and $\tau=2+\kappa^{-2}$, being $\s^2_H$, $\lambda$ and $\kappa$ hyperparameters of the model. The mean and standard deviation of the variances $\s^2_\ell$ are respectively given by $\mathbb{E}[\sigma_{\ell}^2]=\s^2_H e^{-\lambda(\ell-1)}$ and $\mathrm{Std}[\sigma_{\ell}^2]=\kappa \mathbb{E}[\sigma_{\ell}^2]$. The choice of this particular prior is based on the assumption that the channel coefficients $\mathbf{h}_m^\ell$ are \emph{a priori} expected to decay with the index $\ell$, since they model the multipath effect. However, if the data contains enough evidence against this assumption, the posterior distribution will assign higher probability mass to larger values of $\s^2_\ell$.

We depict the corresponding graphical model in Figure~\ref{ch4:fig:graphicalModelMIMO}, with a value of $L=2$.


\section{Inference}\label{sec:inference}

One of the main challenges in Bayesian probabilistic models is posterior inference, which involves the computation of the posterior distribution over the hidden variables in the model given the data. In most models of interest, including {BNP} models, the posterior distribution cannot be obtained in closed form, and an approximate inference algorithm is used instead. In many BNP time series models, inference is carried out via MCMC methods \cite{Robert2005}. In particular, for the IFHMM \cite{VanGael2009}, typical approaches rely on a blocked Gibbs sampling algorithm that alternates between sampling the number of parallel chains and the global variables (emission parameters and transition probabilities) conditioned on the current value of matrices $\Sb$ and $\Xb$, and sampling matrices $\Sb$ and $\Xb$ conditioned on the current value of the global variables. We follow a similar algorithm, which proceeds iteratively as follows:
\begin{itemize}
\item \textbf{Step 1:} Add $M_{\textrm{new}}$ new inactive chains\footnote{An inactive chain is a chain in which all elements $\stm=0$.} (FSMs) using an auxiliary slice variable and a slice sampling method. In this step, the number of considered parallel chains is increased from its initial value $M_+$ to $M^\ddagger=M_++M_{\textrm{new}}$ (we do not update $M_+$ because the new chains are in the all-zero state).
\item \textbf{Step 2:} Sample the states $\stm$ and the input symbols $\xtm$ of all the considered chains. Compact the representation by removing those FSMs that remain inactive in the entire observation period, consequently updating $M_+$.
\item \textbf{Step 3:} Sample the global variables, i.e., the transition probabilities $a^m$ and $b^m$ and the observation parameters $\mathbf{h}_m^\ell$ for each active chain ($m=1,\ldots,M_+$), as well as the variances $\s^2_\ell$.
\end{itemize}

In \textbf{Step 1}, we follow the slice sampling scheme for inference in {BNP} models based on the {IBP} \cite{Teh2007,VanGael2009}, which effectively transforms the model into a finite factorial model with $M^\ddagger=M_++M_{\textrm{new}}$ parallel chains. We first sample an auxiliary slice variable $\slice$, which is distributed as
\begin{equation}\label{ch4:eq:p_slice}
\slice|\Sb,\{a^{(m)}\}_{m=1}^{M_+} \sim \mathrm{Uniform}\left(0,a_{\textrm{min}} \right),
\end{equation}
where $a_{\textrm{min}}=\min_{m:\exists t,s_{tm}\neq 0} a^{(m)}$, and we can replace the uniform distribution with a more flexible scaled beta distribution. Then, starting from $m=M_{+}+1$, new variables $a^{(m)}$ are iteratively sampled from
\begin{equation}\label{ch4:eq:new_am}
\begin{split}
& p(a^{(m)}|a^{(m-1)}) \propto \exp\left( \alpha\sumd_{t=1}^{T}\fracd{1}{t}(1-a^{(m)})^t\right)\\
& \quad \times (a^{(m)})^{\alpha-1}(1-a^{(m)})^{T}\mathbb{I}(0\leq a^{(m)}\leq a^{(m-1)}),
\end{split}
\end{equation}
with $a^{(M_+)}=a_{\textrm{min}}$, until the resulting value is less than the slice variable, i.e., until $a^{(m)}<\slice$. Since Eq.~\ref{ch4:eq:new_am} is log-concave in $\log a^{(m)}$ \cite{Teh2007}, we can apply adaptive rejection sampling (ARS) \cite{Gilks1992}. Let $M_{\textrm{new}}$ be the number of new variables $a^{(m)}$ that are greater than the slice variable. If $M_{\textrm{new}}>0$, then we expand the representation of matrices $\Sb$ and $\Xb$ by adding $M_{\textrm{new}}$ zero columns, and we sample the corresponding per-chain global variables (i.e., $\mathbf{h}_{m}^\ell$ and $b^{m}$) from the prior, given in Eqs.~\ref{ch4:eq:p_hm} and \ref{eq:stick_bm}, respectively.

\textbf{Step 2} consists in sampling the elements of the matrices $\Sb$ and $\Xb$ given the current value of $M^\ddagger$ and the global variables. In this step, several approaches can be taken. A na\"{i}ve Gibbs sampling algorithm that sequentially samples each element $\xtm$ (jointly with $\stm$) is simple and computationally efficient, but it presents poor mixing properties due to the strong couplings between successive time steps \cite{Scott2002,VanGael2009}. An alternative to Gibbs sampling typically applied in factorial HMMs is blocked sampling, which sequentially samples each parallel chain, conditioned on the current value of the remaining ones. This approach requires a forward-filtering backward-sampling (FFBS) sweep in each of the chains, which yields runtime complexity of $\Ocal(TM^\ddagger|\Xcal|^{L+1})$ for our IFFSM model. The exponential dependency on $L$ makes this method computationally intractable.
In order to address this limitation of the FFBS approach, we propose to jointly sample matrices $\Sb$ and $\Xb$ using PGAS, an algorithm recently developed for inference in state-space models and non-Markovian latent variable models \cite{Lindsten2014}. The runtime complexity of this algorithm scales as $\Ocal(PTM^\ddagger L^2)$, where $P$ is the number of particles used for the PGAS kernel. Details on the {PGAS} approach are given in Section~\ref{ch4:subsec:pgas}.


After running {PGAS}, we remove those chains that remain inactive in the whole observation period. This implies removing some columns of $\Sb$ and $\Xb$ as well as the corresponding variables $\mathbf{h}_m^\ell$, $a^m$ and $b^m$, and updating $M_+$.

In \textbf{Step 3}, we sample the global variables in the model from their complete conditional distributions.\footnote{The complete conditional is the conditional distribution of a hidden variable, given the observations and the rest of hidden variables.} The complete conditional distribution over the transition probabilities $a^m$ under the semi-ordered stick-breaking construction \cite{Teh2007} is 
\begin{equation}
a^m|\Sb \sim \textrm{Beta}\left(n_{01}^m,1+n_{00}^m \right),
\end{equation}
being $n_{ij}^m$ the number of transitions from state $i$ to state $j$ in the $m$-th column of $\Sb$. For the self-transition probabilities of the active state $b^m$, we have
\begin{equation}
b^m|\Sb \sim \textrm{Beta}\left(\beta_0+n_{11}^m,\beta_1+n_{10}^m \right).
\end{equation}
The complete conditional distributions over the emission parameters $\mathbf{h}_m^\ell$ for all chains $m=1,\ldots,M_+$ and for all taps $\ell=1,\ldots,L$ are given by complex Gaussians of the form
\begin{equation}\label{ch4:eq:post_hm}
\mathbf{h}^{(d)}|\Yb,\Xb,\{\s^2_\ell\} \sim \Ccal \Ncal \left(\mub^{(d)}_{\textrm{POST}}, \bm{\Gamma}_{\textrm{POST}}, \mathbf{0}\right),
\end{equation}
for $d=1,\ldots,D$, where $\Yb$ is the $T\times D$ matrix containing all vectors $\yt$. Here, we have defined for notational simplicity $\mathbf{h}^{(d)}$ as the vector that contains the $d$-th component of vectors $\mathbf{h}_m^\ell$ for all $m$ and $\ell$, as given by
\begin{equation}
\mathbf{h}^{(d)} = \left[(\mathbf{h}_1^1)_d,\ldots,(\mathbf{h}_1^L)_d,\ldots,
									(\mathbf{h}_{M_+}^1)_d,\ldots,(\mathbf{h}_{M_+}^L)_d
 \right]^\top.
\end{equation}
We can write the parameters $\mub^{(d)}_{\textrm{POST}}$ and $\bm{\Gamma}_{\textrm{POST}}$ as follows. We first define the extended matrix $\Xb^{\textrm{ext}}=\left[ \Xb^{(1)},\ldots,\Xb^{(M_+)} \right]$ of size $T\times LM_+$, with
\begin{equation}
\Xb^{(m)} = \left[ \begin{array}{ccccc}
							x_{1m} & 0 & 0 & \cdots & 0 \\	
							x_{2m} & x_{1m} & 0 & \cdots & 0 \\	
							x_{3m} & x_{2m} & x_{1m} & \cdots & 0 \\	
							\vdots & \vdots & \vdots & \ddots & \vdots \\
							x_{Tm} & x_{(T-1)m} & x_{(T-2)m} & \cdots & x_{(T-L+1)m} \\	
							\end{array}
						     \right],
\end{equation}
$\Sigmab$ as the $L\times L$ diagonal matrix containing all variables $\s^2_\ell$, and $\mathbf{y}^{(d)}$ as the $T$-dimensional vector containing the $d$-th element of each observation $\yt$. Then, the posterior parameters in Eq.~\ref{ch4:eq:post_hm} are given by
\begin{equation}
\bm{\Gamma}_{\textrm{POST}} = \left( (\Ib_{M_+}\otimes \Sigmab)^{-1} +\frac{1}{\s^2_y}(\Xb^\textrm{ext})^\Htop \Xb^\textrm{ext}\right)^{-1}
\end{equation}
and
\begin{equation}
\mub^{(d)}_{\textrm{POST}} = \frac{1}{\s^2_y} \bm{\Gamma}_{\textrm{POST}} (\Xb^\textrm{ext})^\Htop \mathbf{y}^{(d)}, 
\end{equation}
being $(\cdot)^\Htop$ the conjugate transpose, `$\otimes$' the Kronecker product, and $\Ib_{M_+}$ the identity matrix of size $M_+$.

Regarding the complete conditionals of the variances $\s^2_\ell$, they are given by inverse-gamma distributions of the form
\begin{equation}
\begin{split}
& p(\s^2_{\ell}|\{\mathbf{h}_m^\ell\}_{m=1}^{M_+}) \\
& \propto \left(\frac{1}{\s^2_{\ell}}\right)^{1+\tau+DM_+} \exp\left\{-\frac{\nu_\ell+\sum_{m=1}^{M_+}||\mathbf{h}_m^\ell||_2^2}{\s^2_{\ell}}\right\},
\end{split}
\end{equation}
being $||\mathbf{h}_m^\ell||_2^2$ the squared L$^2$-norm of the vector $\mathbf{h}_m^\ell$.

\subsection{Particle Gibbs with Ancestor Sampling}\label{ch4:subsec:pgas}

We rely on {PGAS} \cite{Lindsten2014} for Step 2 of our inference algorithm, in order to obtain a sample of the matrices $\Sb$ and $\Xb$ (which at this stage of the inference algorithm are truncated to $M^\ddagger$ columns). {PGAS} is a method within the framework of PMCMC \cite{Andrieu2010}, which is a systematic way of combining sequential Monte Carlo (SMC) and {MCMC} to take advantage of the strengths of both techniques.

In {PGAS}, a Markov kernel is constructed by running an SMC sampler in which one particle is set deterministically to a reference input particle. This reference particle corresponds to the output of the previous PGAS iteration (extended to account for the $M_{\textrm{new}}$ new FSMs). After a complete run of the {SMC} algorithm, a new reference trajectory is obtained by selecting one of the particle trajectories with probabilities given by their importance weights. In this way, the resulting Markov kernel leaves its target distribution invariant, regardless of the number of particles used. In contrast to other particle Gibbs with backward simulation methods \cite{Whiteley2010,Lindsten2013}, {PGAS} can also be applied to non-Markovian latent variable models, i.e., models that are not expressed on a state-space form \cite{Lindsten2014,Valera2015nips}.
In this section, we briefly describe the PGAS algorithm and provide the necessary equations for its implementation. 

\begin{figure}[t]
\includegraphics[width=7cm]{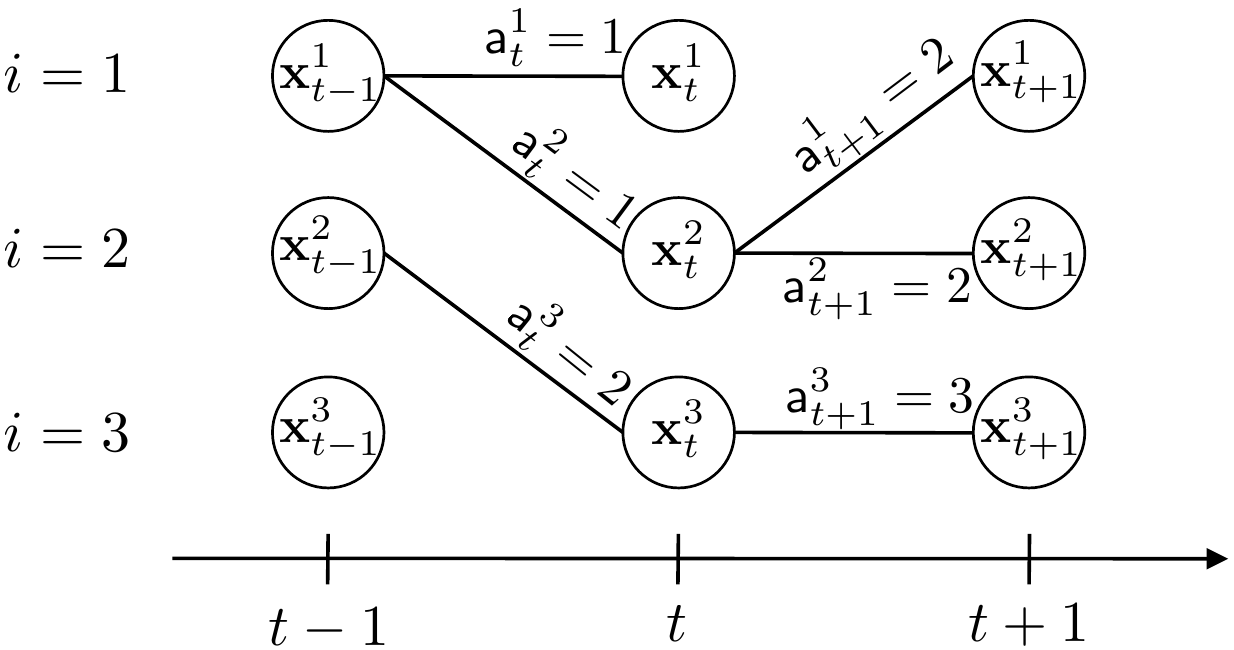}
\centering
\caption[Example of the connection of particles in PGAS]{Example of the connection of particles in PGAS. We represent $P=3$ particles $\mathbf{x}_{\tau}^i$ for $\tau=\{t-1,t,t+1\}$. The index $\mathsf{a}_{\tau}^i$ denotes the ancestor particle of $\mathbf{x}_{\tau}^i$. It can be seen that, e.g., the trajectories $\mathbf{x}_{1:t+1}^1$ and $\mathbf{x}_{1:t+1}^2$ only differ at time instant $t+1$.}
\label{ch4:fig:examplePGAS}
\end{figure}

In {PGAS}, we assume a set of $P$ particles for each time instant $t$, each representing the hidden states $\{\xtm\}_{m=1}^{M^\ddagger}$ (hence, they also represent $\{\stm\}_{m=1}^{M^\ddagger}$). We denote the state of the $i$-th particle at time $t$ by the vector $\xt^i$ of length $M^\ddagger$. Similarly, the input reference particle for each time instant is denoted as $\mathbf{x}_t^\prime$. We also introduce the ancestor indexes $\mathsf{a}_t^i\in\{1,\ldots,P\}$ in order to denote the particle that precedes the $i$-th particle at time $t$. That is, $\mathsf{a}_t^i$ corresponds to the index of the ancestor particle of $\xt^i$. Let also $\mathbf{x}_{1:t}^i$ be the ancestral path of particle $\xt^i$, i.e., the particle trajectory that is recursively defined as
\begin{equation}\label{ch4:eq:pgas_recTrajectory}
\mathbf{x}_{1:t}^i = (\mathbf{x}_{1:t-1}^{\mathsf{a}_t^i},\xt^i).
\end{equation}
Figure~\ref{ch4:fig:examplePGAS} shows an example to clarify the notation.

The machinery inside the PGAS algorithm resembles an ordinary particle filter, with two main differences: one of the particles is deterministically set to the reference input sample, and the ancestor of each particle is randomly chosen and stored during the algorithm execution. 
Algorithm~\ref{alg:pgas} summarizes the procedure. For each time instant $t$, we first generate the ancestor indexes for the first $P-1$ particles according to the importance weights $w_{t-1}^i$. Given these ancestors, the particles are then propagated across time according to a distribution $r_t(\xt|\mathbf{x}_{1:t-1}^{\mathsf{a}_t})$. For simplicity, and dropping the global variables from the notation for conciseness, we assume that
\begin{equation}\label{ch4:eq:pgas_rxt}
\begin{split}
& r_t(\xt|\mathbf{x}_{1:t-1}^{\mathsf{a}_t})
=p(\xt|\mathbf{x}_{t-1}^{\mathsf{a}_t}) \\
& =\prodd_{m=1}^{M^\ddagger} p(\xtm|\stm)p(\stm|s_{(t-1)m}^{\mathsf{a}_t}),
\end{split}
\end{equation}
i.e., particles are propagated as in Figure~\ref{ch4:fig:graphicalModel} using a simple bootstrap proposal kernel, $p(\xtm,\stm|\sttm, \xttm)$. The $P$-th particle is instead deterministically set to the reference particle, $\mathbf{x}_t^P=\mathbf{x}_t^\prime$, whereas the ancestor indexes $\mathsf{a}_t^P$ are sampled according to some weights $\widetilde{w}_{t-1|T}^i$. Indeed, this is a crucial step that vastly improves the mixing properties of the MCMC kernel.

\begin{algorithm}[t]
\caption{Particle Gibbs with ancestor sampling}
\label{alg:pgas}
\begin{algorithmic}[1]
\REQUIRE Reference particle $\mathbf{x}_t^{\prime}$ for $t=1,\ldots,T$, and global variables
\ENSURE Sample $\mathbf{x}_{1:T}^{\mathrm{out}}$ from the PGAS Markov kernel
\STATE Draw $\mathbf{x}_1^i\sim r_1(\mathbf{x}_1)$ for $i=1,\ldots,P-1$ (Eq.~\ref{ch4:eq:pgas_rxt})
\STATE Set $\mathbf{x}_1^P=\mathbf{x}_1^{\prime}$
\STATE Compute the weights $w_1^i$ for $i=1,\ldots,P$ (Eq.~\ref{ch4:eq:pgas_impWeights})
\FOR{$t=2,\ldots,T$}
\STATE Draw $\mathsf{a}_t^i \sim \textrm{Categorical}(w_{t-1}^1,\ldots,w_{t-1}^P)$ for $i=1,\ldots,P-1$
\STATE Compute $\widetilde{w}_{t-1|T}^i$ for $i=1,\ldots,P$ (Eq.~\ref{ch4:eq:pgas_ancWeights})
\STATE Draw $\mathsf{a}_t^P \sim \textrm{Categorical}(\widetilde{w}_{t-1|T}^1,\ldots,\widetilde{w}_{t-1|T}^P)$ 
\STATE Draw $\mathbf{x}_t^i\sim r_t(\mathbf{x}_t|\mathbf{x}_{1:t-1}^{\mathsf{a}_t^i})$ for $i=1,\ldots,P-1$ (Eq.~\ref{ch4:eq:pgas_rxt})
\STATE Set $\mathbf{x}_t^P=\mathbf{x}_t^{\prime}$
\STATE Set $\mathbf{x}_{1:t}^i=(\mathbf{x}_{1:t-1}^{\mathsf{a}_t^i},\mathbf{x}_t^i)$ for $i=1,\ldots,P$ (Eq.~\ref{ch4:eq:pgas_recTrajectory})
\STATE Compute the weights $w_t^i$ for $i=1,\ldots,P$ (Eq.~\ref{ch4:eq:pgas_impWeights})
\ENDFOR
\STATE Draw $k\sim \textrm{Categorical}(w_T^1,\ldots,w_T^P)$ 
\RETURN $\mathbf{x}_{1:T}^{\mathrm{out}}=\mathbf{x}_{1:T}^k$
\end{algorithmic}
\end{algorithm}

We now focus on the computation of the importance weights $w_t^i$ and the ancestor weights $\widetilde{w}_{t-1|T}^i$. For the former, the particles are weighted according to 
\begin{equation}\label{ch4:eq:pgas_impWeights}
\begin{split}
w_t^i & =\frac{p(\mathbf{x}_{1:t}|\mathbf{y}_{1:t})}{p(\mathbf{x}_{1:t-1}|\mathbf{y}_{1:t-1})r_t(\xt|\mathbf{x}_{1:t-1})} \\
& \propto \frac{p(\mathbf{y}_{1:t}|\mathbf{x}_{1:t})p(\mathbf{x}_{1:t})}{p(\mathbf{y}_{1:t-1}|\mathbf{x}_{1:t-1})p(\mathbf{x}_{1:t-1})p(\xt|\mathbf{x}_{t-1})} \\
& = p(\mathbf{y}_{t}|\mathbf{x}_{t-L+1:t}),
\end{split}
\end{equation}
being $\mathbf{y}_{\tau_1:\tau_2}$ the set of observations $\{\yt\}_{t=\tau_1}^{\tau_2}$. We have applied \eqref{ch4:eq:pgas_rxt} to derive this expression. Eq.~\ref{ch4:eq:pgas_impWeights} implies that, in order to obtain the importance weights, it suffices to evaluate the likelihood at time $t$.

The weights $\widetilde{w}_{t-1|T}^i$ used to draw a random ancestor for the reference particle are given by
\begin{equation}\label{ch4:eq:pgas_ancWeights}
\begin{split}
\widetilde{w}_{t-1|T}^i & =w_{t-1}^i\frac{p(\mathbf{x}_{1:t-1}^{i},\mathbf{x}_{t:T}^{\prime}|\mathbf{y}_{1:T})}{p(\mathbf{x}_{1:t-1}^{i}|\mathbf{y}_{1:t-1})} \\
& \propto w_{t-1}^i\frac{p(\mathbf{y}_{1:T}|\mathbf{x}_{1:t-1}^{i},\mathbf{x}_{t:T}^{\prime})p(\mathbf{x}_{1:t-1}^{i},\mathbf{x}_{t:T}^{\prime})}{p(\mathbf{y}_{1:t-1}|\mathbf{x}_{1:t-1}^{i})p(\mathbf{x}_{1:t-1}^{i})} \\
& \propto w_{t-1}^i p(\mathbf{x}_t^{\prime}|\mathbf{x}_{t-1}^i) \prodd_{\tau=t}^{t+L-2} p(\mathbf{y}_{\tau}|\mathbf{x}_{1:t-1}^i,\mathbf{x}_{t:T}^{\prime}).
\end{split}
\end{equation}
In order to obtain this expression, we have made use of the Markov property of the model, and we have also ignored factors that do not depend on the particle index $i$. Note that the transition probability $p(\mathbf{x}_t|\mathbf{x}_{t-1})$ factorizes across the parallel chains of the factorial model, as given in \eqref{ch4:eq:pgas_rxt}. We also note that, for memoryless models (i.e., $L=1$), Eq.~\ref{ch4:eq:pgas_ancWeights} can be simplified, since the product in the last term is not present and, therefore, $\widetilde{w}_{t-1|T}^i \propto w_{t-1}^i p(\mathbf{x}_t^{\prime}|\mathbf{x}_{t-1}^i)$.

\subsection{Computational Complexity}
{For any channel length $L$, the resulting complexity of each iteration of the algorithm scales as $\Ocal(PTM^\ddagger L^2)$. This is because the most expensive step is the computation of the weights $\widetilde{w}_{t-1|T}^i$ in }\eqref{ch4:eq:pgas_ancWeights}{ for $i=1,\ldots,P$, which has complexity scaling as $\Ocal(PM^\ddagger L^2)$. This computation needs to be performed for each time instant $t=1,\ldots,T$, and hence the resulting overall complexity.}



\begin{figure}[h]
\centering
\subfloat[Inferred $M_+$.]{\includegraphics[width=.25\textwidth]{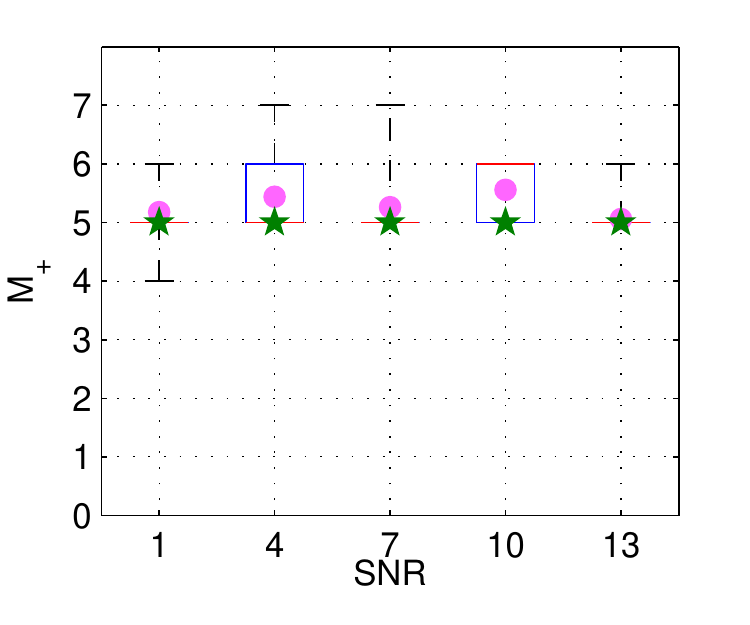}} 
\subfloat[Recovered transmitters.]{\includegraphics[width=.25\textwidth]{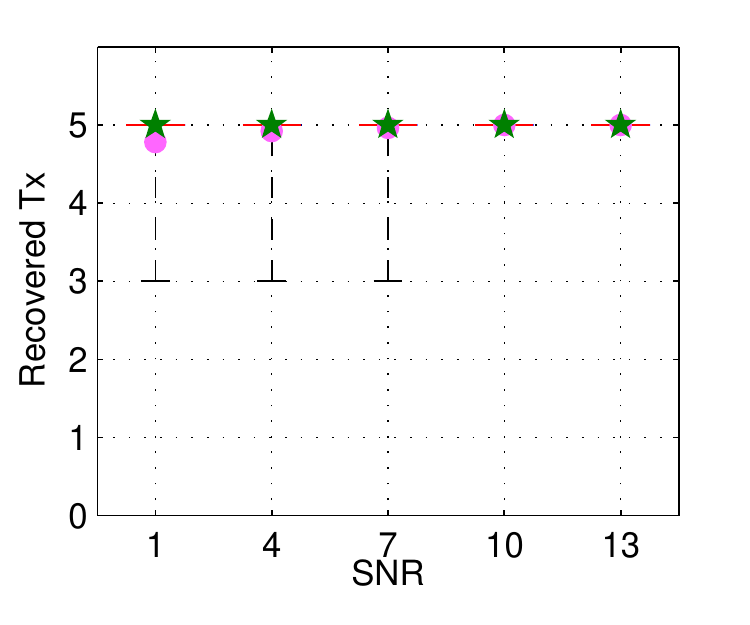}} \\ \vspace*{-10pt} 
\subfloat[ADER.]{\includegraphics[width=.25\textwidth]{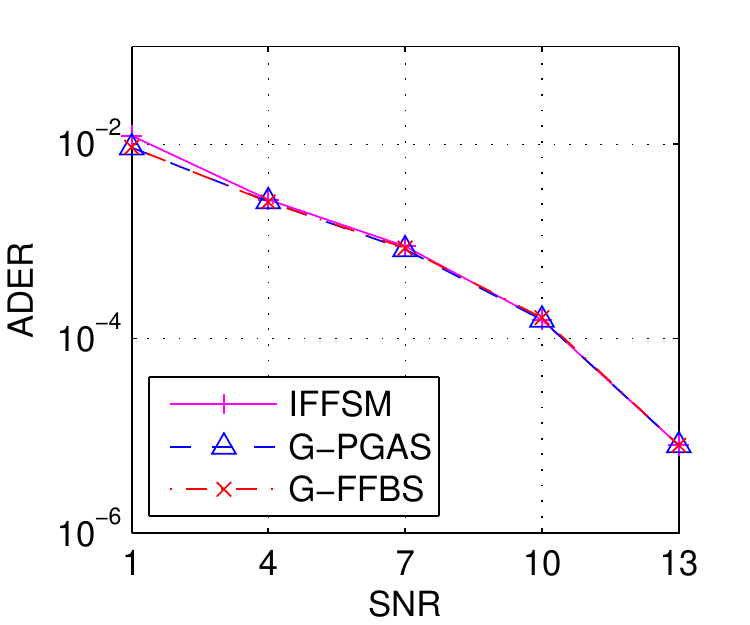}}
\subfloat[SER.]{\includegraphics[width=.25\textwidth]{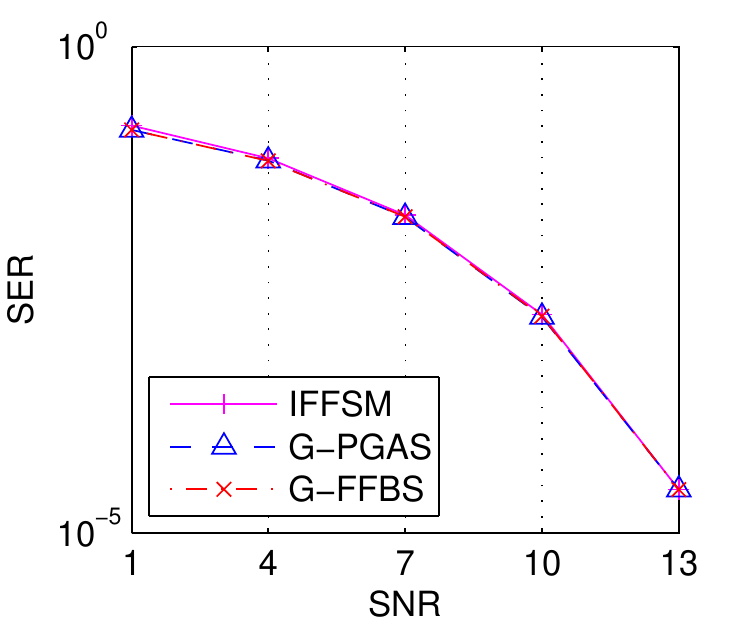}} \\ \vspace*{-10pt} 
\subfloat[MSE.]{\includegraphics[width=.25\textwidth]{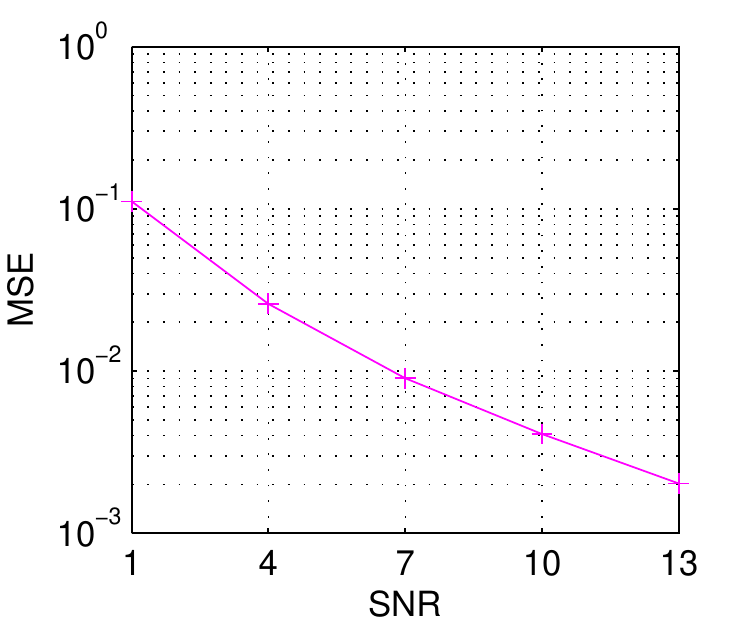}} 
\caption{{Results for different SNRs ($L=1$).}}\label{ch6:fig:resultsSNR}  \vspace*{-10pt}
\end{figure}

\section{Experiments}\label{sec:experiments}

In this section, we apply the {IFFSM} model in Section~\ref{sec:model} to the problem of user activity detection and blind channel estimation. We use two different channel models, namely, a standard multipath Rayleigh fading channel model in Section~\ref{subsec:synthetic} and a ray-tracing model in Section~\ref{subsec:realData}.

\subsection{Multipath Rayleigh Fading}\label{subsec:synthetic}

In LTE, when the user equipment (UE) wants to transmit information, it needs to ask for resources in the random access channel (RACH) \cite{Sesia2011}. This is the first of four messages between the enhanced Node B (eNB), i.e., the network access point, and the UE before the actual transmission of information starts. The UE selects one of $64$ random sequences with $839$ symbols and uses $72$ subcarriers ($1.08$ MHz) to transmit the sequence in $0.8$ ms in a $1$-ms sub-frame ($0.2$ ms are used for cyclic prefix and time guard band). In the typical configuration, there is a RACH subframe every $10$-ms frame. If two UEs transmit the same preamble sequence at the same physical RACH, one of them (or both) would not get a response from the network and it (they) would need to resend the RACH preamble, although in most cases this channel goes unused. 

In our simulations, we assume that instead of a RACH preamble asking for resources, our IoT devices send a $500$-symbol sequence with the payload that they want to transmit for $0.5$ ms, and the remaining $0.5$ ms are reserved for time guard band. The timing of the IoT might not be as accurate as typical cellphones, and hence we allow for much larger guard band. We assume that the devices can start their transmission at any time in the first half of the physical RACH, and we also assume $5$ active users in each physical RACH channel interval. We consider a Rayleigh {AWGN} channel, i.e., the channel coefficients and the noise are circularly-symmetric complex Gaussian distributed with zero mean, where the covariances matrices are $\sigma_\ell^2 \mathbf{I}$ and $\sigma_y^2 \mathbf{I}$, respectively.
We simulate a base scenario with $D=20$ receiving antennas, quadrature phase-shift keying (QPSK) modulation {(the constellation is normalized to yield unit energy)}
, and $\sigma_y^2=2$. Using this base configuration, we vary one of the parameters while holding the rest fixed. 

We set the hyperparameters of the IFFSM as $\sigma_H^2=1$, $\lambda=0.5$, $\kappa=1$, $\alpha=1$, $\beta_0=2$ and $\beta_1=0.1$. The choice of $\beta_0$ and $\beta_1$ is based on the fact that we expect the active Markov chains to remain active and, therefore, the transition probabilities from active to active $b^m$, which are $\textrm{Beta}(\beta_0,\beta_1)$ distributed, are \emph{a priori} expected to be of the order of one over a few hundred bits. In order to avoid getting trapped in local modes of the posterior, we use a tempering procedure. We first add artificial noise so that the resulting $\sigma_y^2=10^{1.2}$ and after each iteration we reduce this noise by a factor of $0.9995$, until there is no artificial noise left. After that, we run additional iterations to favor exploitation.

\begin{figure}[t]
\centering
\subfloat[Inferred $M_+$.]{\includegraphics[width=.25\textwidth]{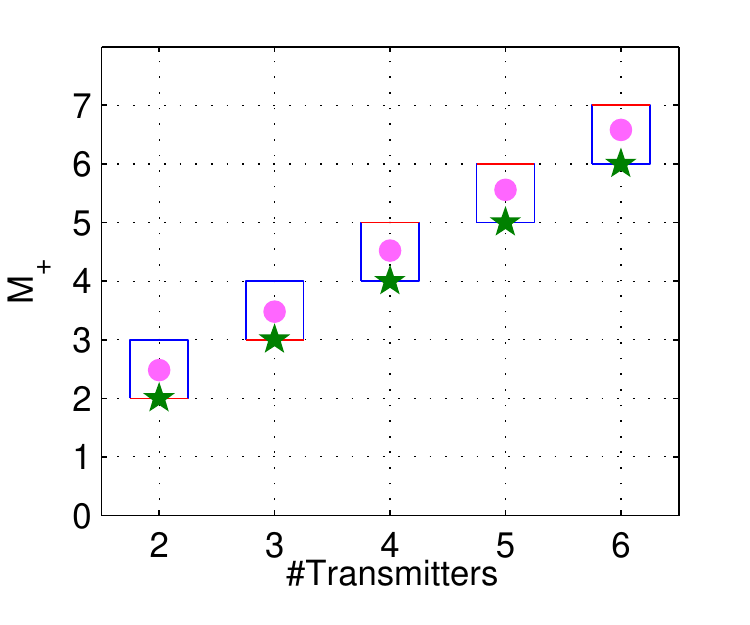}}
\subfloat[Recovered transmitters.]{\includegraphics[width=.25\textwidth]{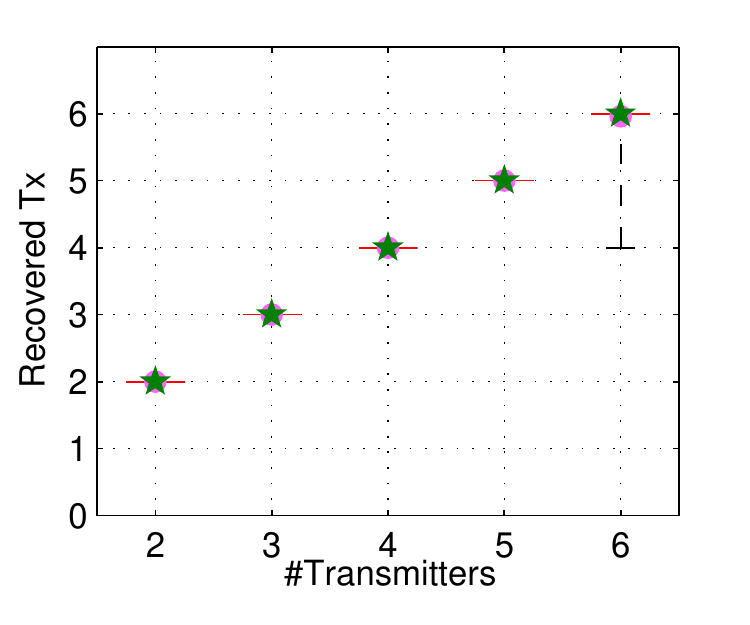}} \\ \vspace*{-10pt}
\subfloat[ADER.]{\includegraphics[width=.25\textwidth]{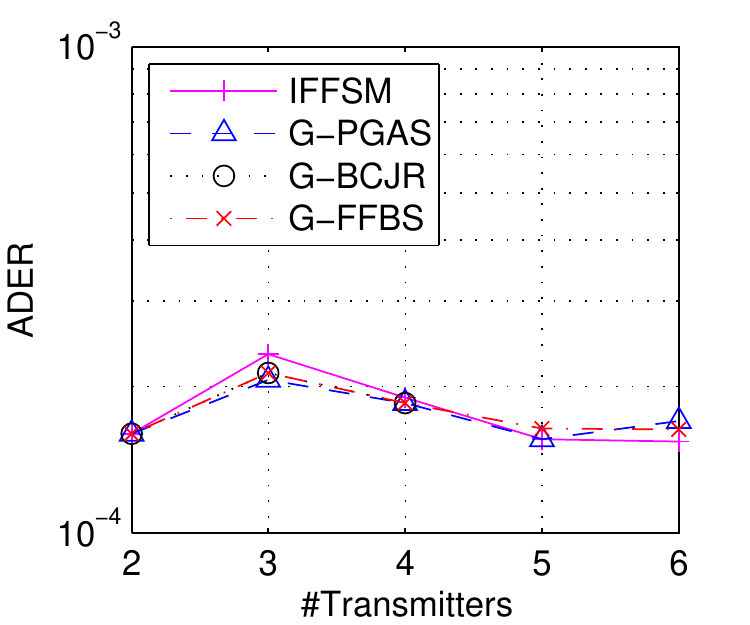}}
\subfloat[SER.]{\includegraphics[width=.25\textwidth]{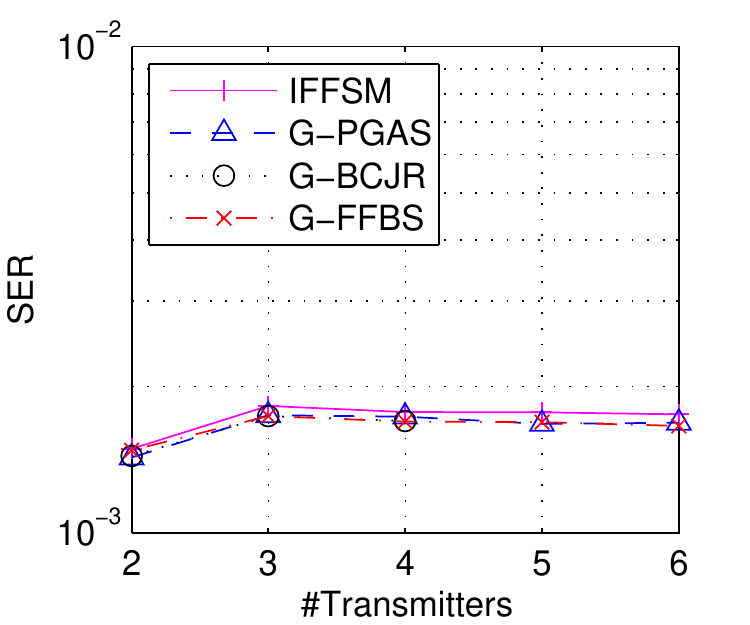}} \\ \vspace*{-10pt}
\subfloat[MSE.]{\includegraphics[width=.25\textwidth]{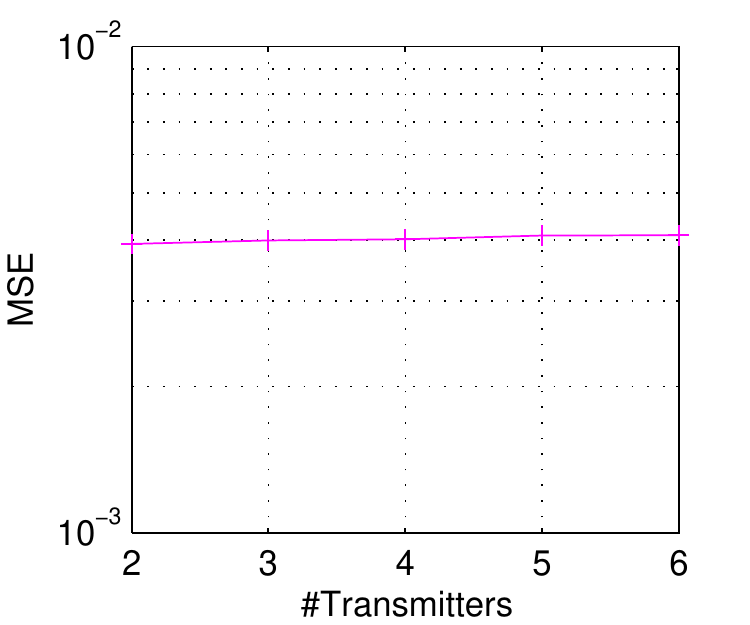}}
\caption{Results for different number of transmitters ($L=1$).}\label{ch6:fig:resultsNt} \vspace*{-10pt}
\end{figure}

\xhdr{Evaluation} 
For the recovered transmitters,\footnote{Our procedure is fully blind, so it might return spurious transmitters or it might mix two or more transmitters in a single chain. We discard those chains in our evaluations.} we evaluate the performance in terms of the activity detection error rate (ADER), the symbol error rate (SER) and the mean squared error (MSE) of the channel coefficient estimates. The {ADER} is the probability of detecting activity (inactivity) in a transmitter while that transmitter is actually inactive (active). When computing the {SER}, an error is computed at time $t$ whenever the estimated symbol for a transmitter differs from the actual transmitted symbol, considering that the transmitted symbol while inactive is $\xtm=0$. The {MSE} for each transmitter is
\begin{equation}\label{ch6:eq:mseBis}
\mathrm{MSE}_m = \frac{1}{LD} \sumd_{d,\ell} \left|\left|(\mathbf{h}^{\ell}_{m})_d - (\hat{\mathbf{h}}^{\ell}_{m})_d\right|\right|^2,
\end{equation} 
being $\hat{\mathbf{h}}^{\ell}_{m}$ the inferred channel coefficients. 

We compare our approach (denoted by {IFFSM} in the plots) with three genie-aided methods which have perfect knowledge of the true number of transmitters and channel coefficients.\footnote{For the genie-aided methods, we use $a^m=0.998$ and $b^m=0.002$.} In particular, we run: (i) The {PGAS} algorithm that we use in Step 2 of our inference algorithm (G-PGAS), (ii) the {FFBS} algorithm over the equivalent factorial {HMM} with state space cardinality $| \Acal \bigcup \{0\} |^{L}$ (G-FFBS), and (iii) the optimum BCJR algorithm \cite{Bahl1974}, over an equivalent single {HMM} with a number of states equal to $| \Acal \bigcup \{0\} |^{LN_t}$, being $N_t$ the true number of transmitters (G-BCJR). Due to their complexity, we only run the BCJR algorithm in scenarios with $| \Acal \bigcup \{0\} |^{2LN_t}\leq 10^6$, and the {FFBS} in scenarios with $| \Acal \bigcup \{0\} |^{2L}\leq 10^6$.

\begin{figure}[t]
\centering
\subfloat[Inferred $M_+$.]{\includegraphics[width=.25\textwidth]{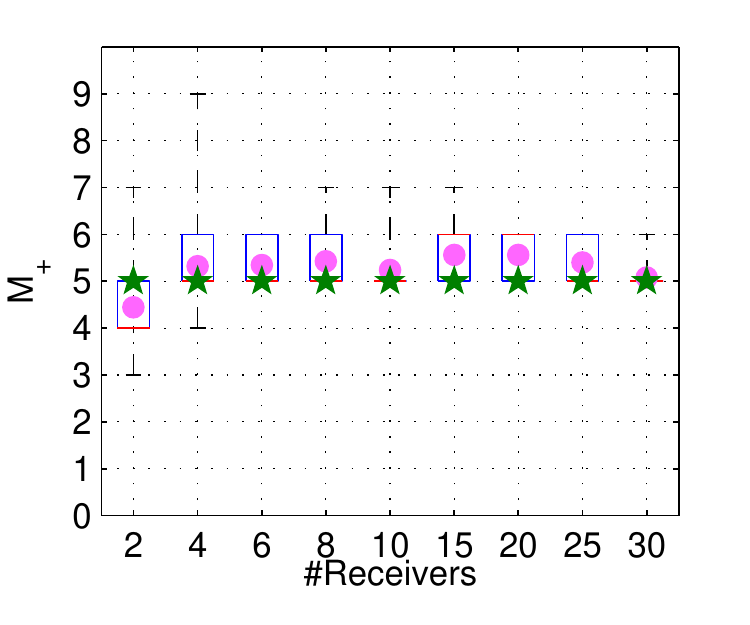}}
\subfloat[Recovered transmitters.]{\includegraphics[width=.25\textwidth]{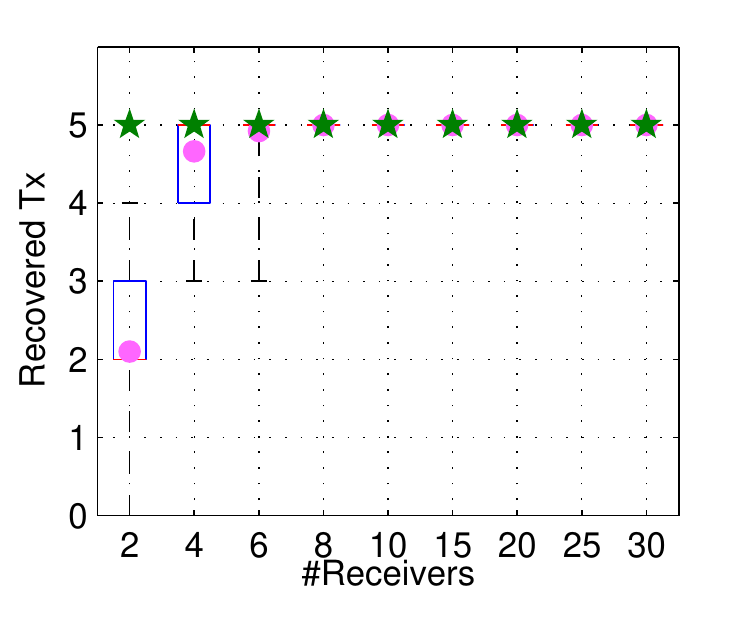}} \\ \vspace*{-10pt}
\subfloat[ADER.]{\includegraphics[width=.25\textwidth]{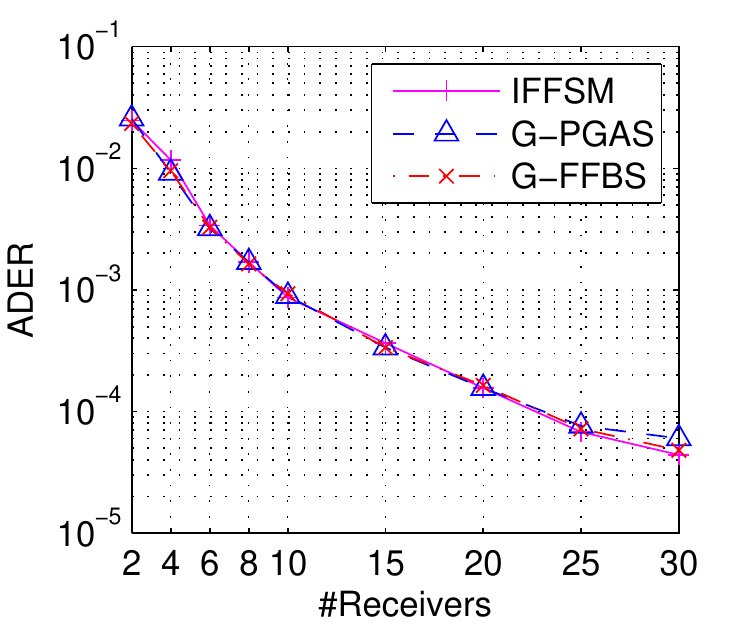}}
\subfloat[SER.]{\includegraphics[width=.25\textwidth]{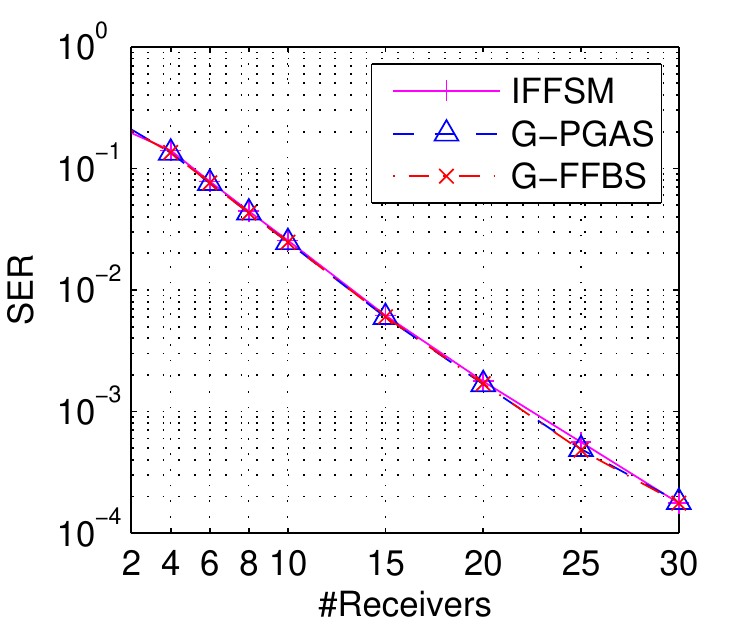}} \\ \vspace*{-10pt}
\subfloat[MSE.]{\includegraphics[width=.25\textwidth]{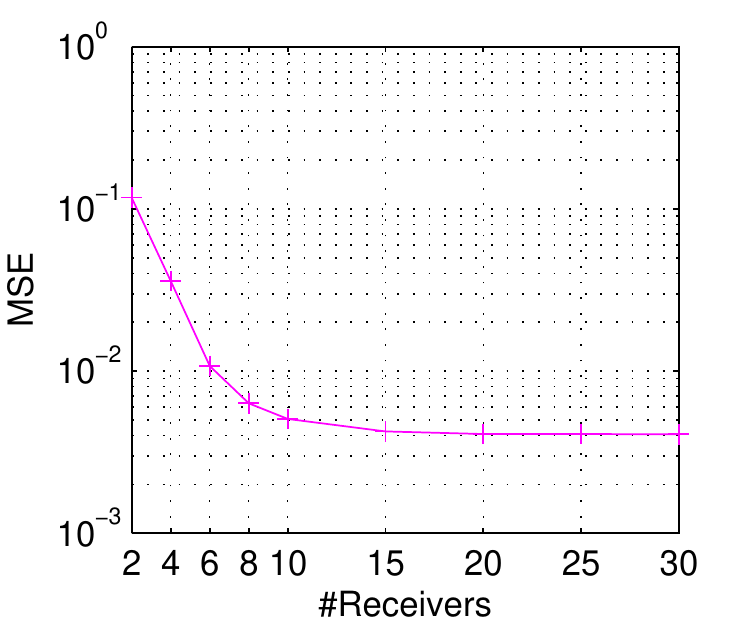}}
\caption{Results for different number of receiving antennas ($L=1$).}\label{ch6:fig:resultsNr} \vspace*{-10pt}
\end{figure}

For each considered scenario, we run $50$ independent simulations, each with different simulated data. We run $20000$ iterations of our inference algorithm, and obtain the inferred symbols $\hat{x}_{tm}$ as the component-wise \textit{maximum a posteriori} (MAP) solution over the last $2000$ samples. The estimates of the channel coefficients $\hat{\mathbf{h}}_m^\ell$ are then obtained as the {MAP} solution, conditioned on the data and the inferred symbols $\hat{x}_{tm}$. For the BCJR algorithm, we obtain the symbol estimates according to the component-wise {MAP} solution for each transmitter $m$ and each instant $t$. For the genie-aided {PGAS} and {FFBS} methods, we follow a similar approach by running the algorithms for $10000$ iterations and considering the last $2000$ samples to obtain the symbol estimates. Unless otherwise specified, we use $P=3000$ particles for {PGAS}.

\begin{figure}[t]
\centering
\subfloat[Inferred $M_+$.]{\includegraphics[width=.25\textwidth]{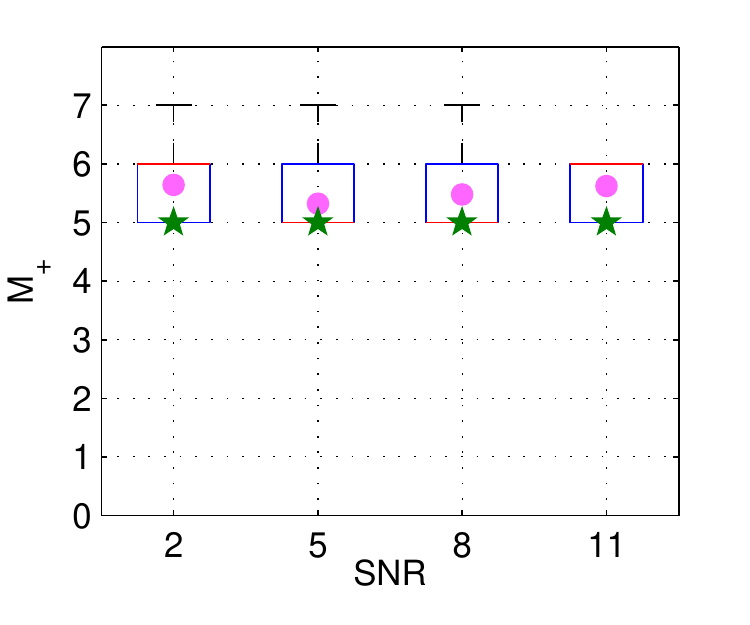}}
\subfloat[Recovered transmitters.]{\includegraphics[width=.25\textwidth]{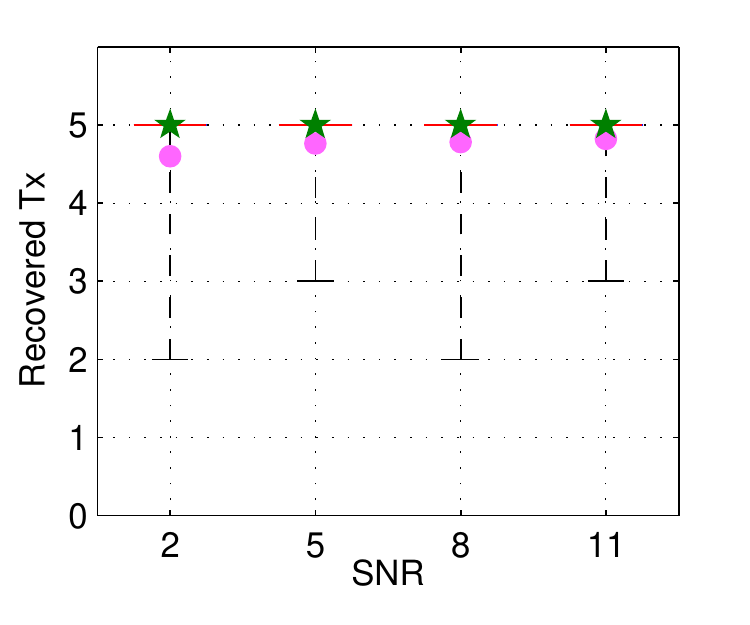}} \\ \vspace*{-10pt}
\subfloat[ADER.]{\includegraphics[width=.25\textwidth]{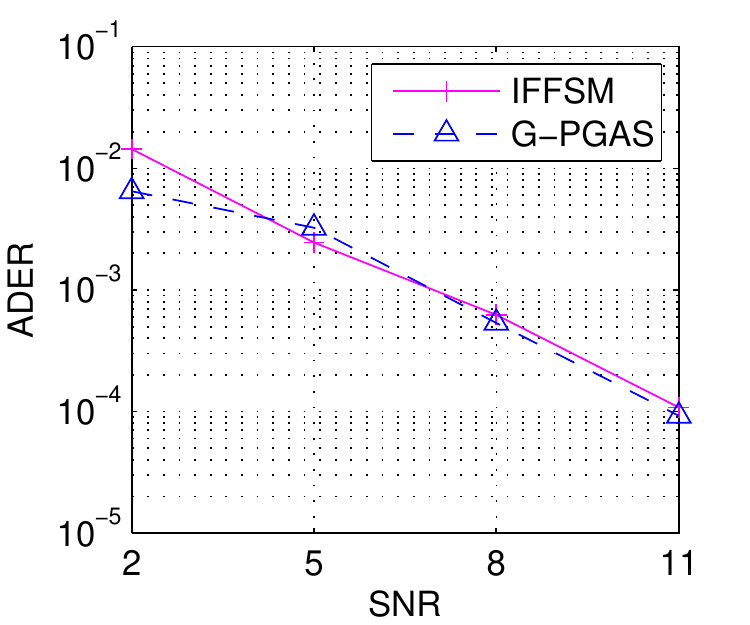}}
\subfloat[SER.]{\includegraphics[width=.25\textwidth]{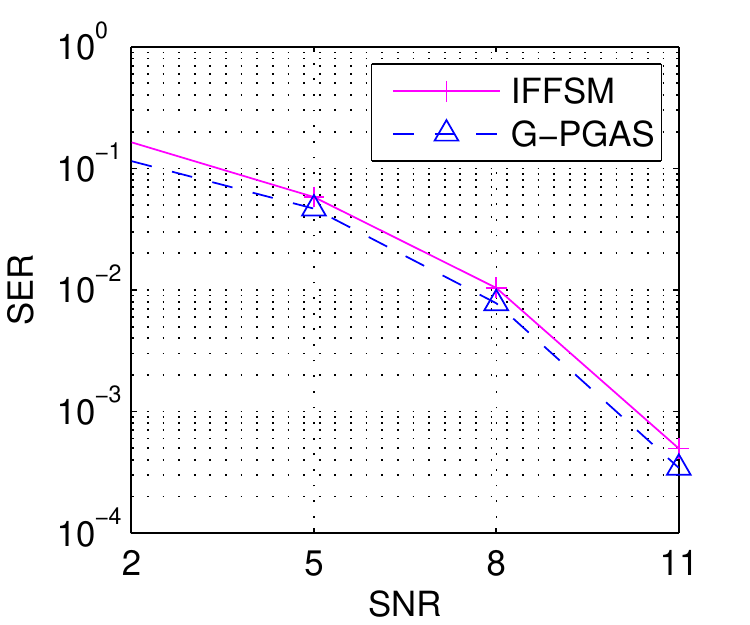}} \\ \vspace*{-10pt}
\subfloat[MSE.]{\includegraphics[width=.25\textwidth]{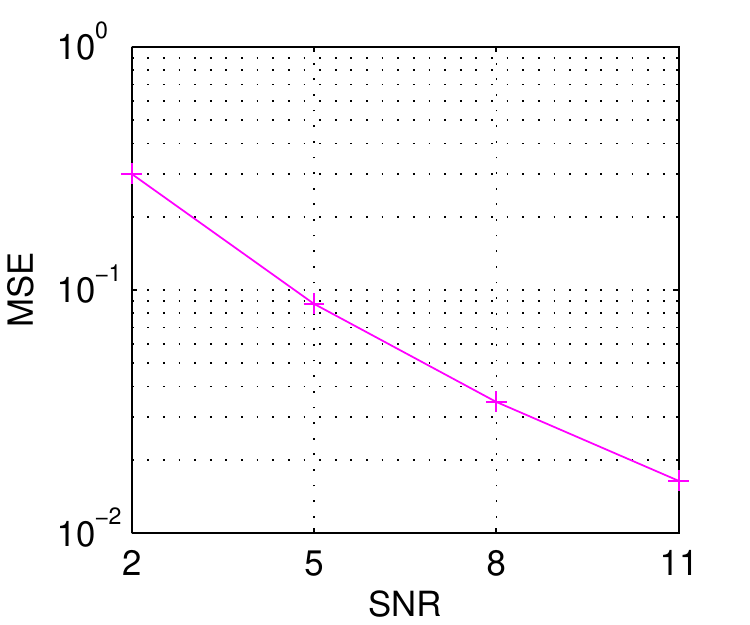}}
\caption{{Results for different SNRs ($L=5$).}}\label{ch6:fig:resultsSNR_L5} \vspace*{-10pt}
\end{figure}

\xhdr{Results with perfect knowledge of the channel memory}
We first evaluate the performance of our model and inference procedure assuming that the memory of the channel is accurately known. We initially consider memoryless channels, i.e., with memory length $L=1$. 
Figure~\ref{ch6:fig:resultsSNR} shows the results when the {noise variance} varies from $\sigma_y^2=10^{1.2}$ to $\sigma_y^2=1$ {(SNR from 1dB to 13dB)}.\footnote{We obtain the per-user SNR (in dB) as $10\log_{10}\left(\frac{DL}{\sigma_y^2}\right)$.} Specifically, we first show a box-plot representation\footnote{We depict the 25-th, 50-th and 75-th percentiles in the standard format, as well as the most extreme values. Moreover, the mean value is represented with a pink circle, and the true number of transmitters is represented with a green star.} of the inferred number of transmitters $M_+$, and also a box-plot representation of the number of recovered transmitters (i.e., how many of the true transmitters we are able to recover). For the recovered transmitters, we additionally show the {ADER}, the {SER}, and the {MSE}. As expected, the performance improves with the {SNR}. For low values of the {SNR}, transmitters are more likely to be masked by the noise and, therefore, on average we recover a slightly lower number of transmitters. We also observe that the performance (in terms of {ADER} and {SER}) of the proposed {IFHMM} is very similar to {the genie-aided methods, which have} perfect knowledge of the number of transmitters and channel coefficients.

Figure~\ref{ch6:fig:resultsNt} shows the results when the true number of transmitters ranges from 2 to 6. Although the number of parameters to be estimated grows with the number of transmitters, we observe that the performance is approximately constant. The {IFHMM} recovers all the transmitters in nearly all the simulations, with performance similar to the genie-aided methods.

\begin{figure}[t]
\centering
\subfloat[Inferred $M_+$.]{\includegraphics[width=.25\textwidth]{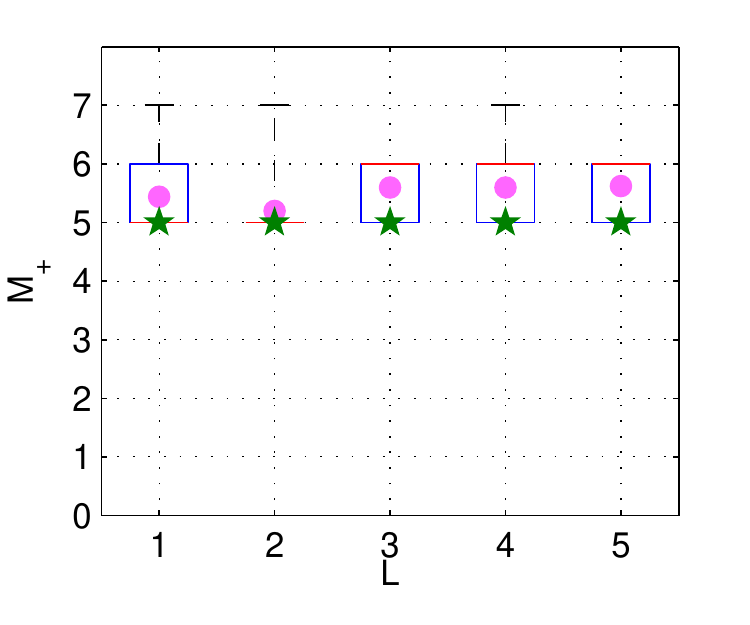}}
\subfloat[Recovered transmitters.]{\includegraphics[width=.25\textwidth]{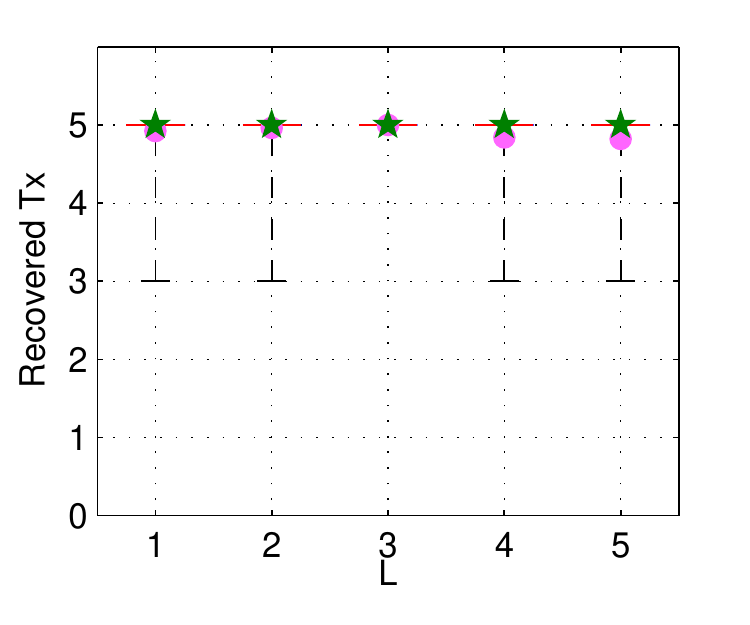}} \\ \vspace*{-10pt}
\subfloat[ADER.]{\includegraphics[width=.25\textwidth]{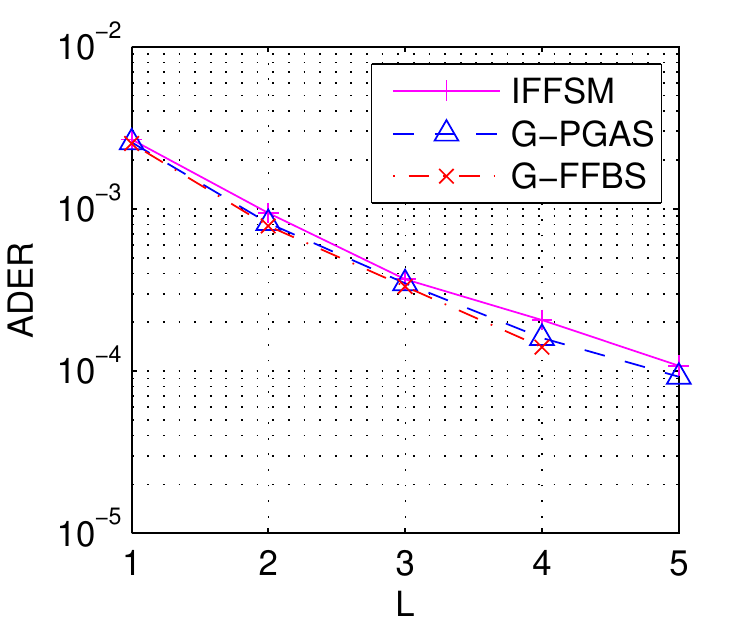}}
\subfloat[SER.]{\includegraphics[width=.25\textwidth]{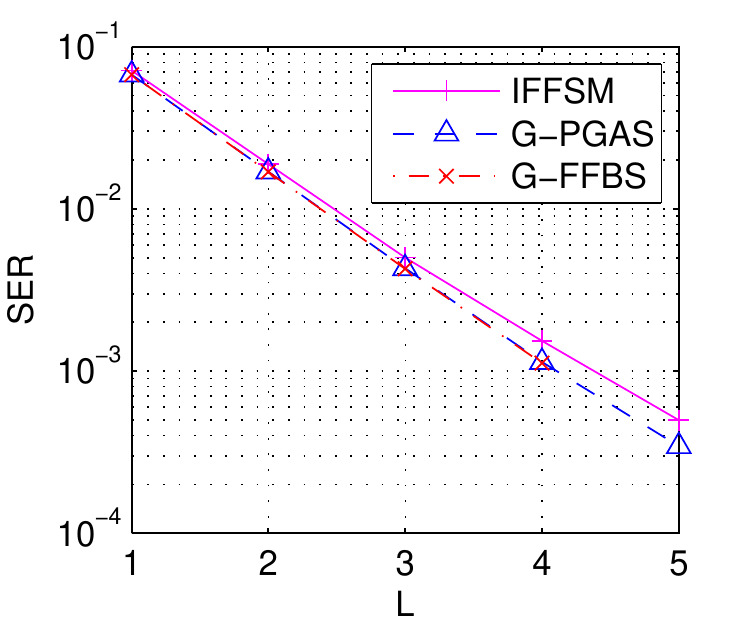}} \\ \vspace*{-10pt}
\subfloat[MSE.]{\includegraphics[width=.25\textwidth]{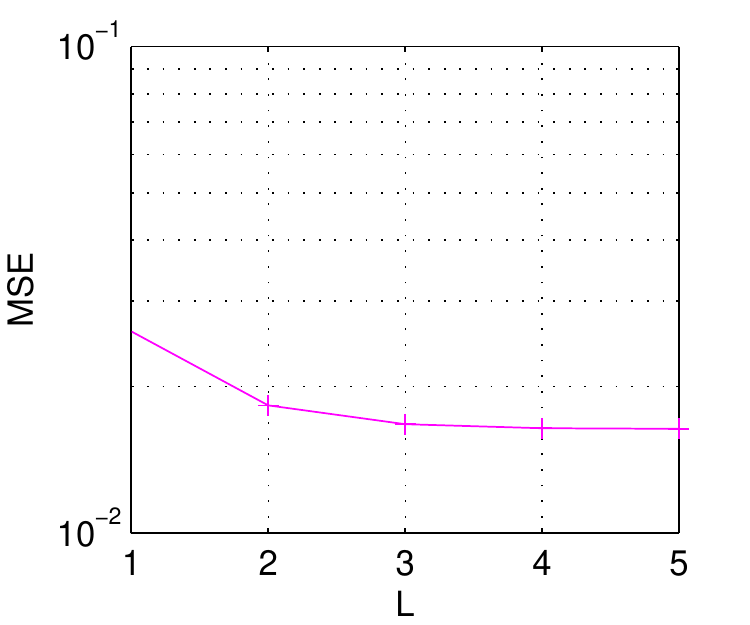}}
\caption{Results for different values of $L$.}\label{ch6:fig:results_Ltrue} \vspace*{-10pt}
\end{figure}

Figure~\ref{ch6:fig:resultsNr} shows the results when the number of receiving antennas varies from 2 to 30. In this figure, we observe that we need at least 6 receivers in order to properly recover the messages of all the transmitters. We see no further improvement in detectability with more than 8 receivers. As expected, the performance in terms of {ADER} and {SER} improves when the number of receiving antennas increases, as the diversity in the observations helps to recover the transmitted symbols and channel coefficients. This behavior is similar to the obtained by the genie-aided {PGAS} and {FFBS}, as shown in this figure. Note that the {MSE} curve flattens after approximately $15$ receivers, as it reaches the threshold imposed by the noise level.

%

We now evaluate the performance of our model and inference procedure for frequency-selective channels, i.e., considering $L>1$. Figure~\ref{ch6:fig:resultsSNR_L5} shows the results when the {noise variance} varies from $\sigma_y^2=10^{1.8}$ to $\sigma_y^2=10^{0.9}$ {(SNR from 2dB to 11dB)}, considering $L=5$ to generate the data. We use the true value of the channel length $L$ for inference. In the figure, we show the {ADER}, the {SER}, the {MSE}, a box-plot representation of the inferred number of transmitters $M_+$, and also a box-plot representation of the number of recovered transmitters. As in the memoryless case, the performance improves with the {SNR}. However, the values of the resulting SER and ADER are much lower than in the memoryless case for a fixed value of the {noise variance}. For instance, the SER in Figure~\ref{ch6:fig:resultsSNR} for $\sigma_y^2=10^{1.2}$ {(SNR=$1$dB)} is one order of magnitude above the reported SER in Figure~\ref{ch6:fig:resultsSNR_L5} for the same value of the {noise variance (SNR=$8$dB)}. This is a sensible result, because the channel memory adds more redundancy in the observed sequence{, and therefore the resulting SNR is higher for a fixed value of the noise variance}. Our inference algorithm is able to exploit such redundancy to better estimate the transmitted symbols, despite the fact that more channel coefficients need to be estimated. Note that the performance in terms of {ADER} and {SER} is similar to the genie-aided {PGAS}-based method.

\begin{figure}[t]
\centering
\subfloat[Inferred $M_+$.]{\includegraphics[width=.25\textwidth]{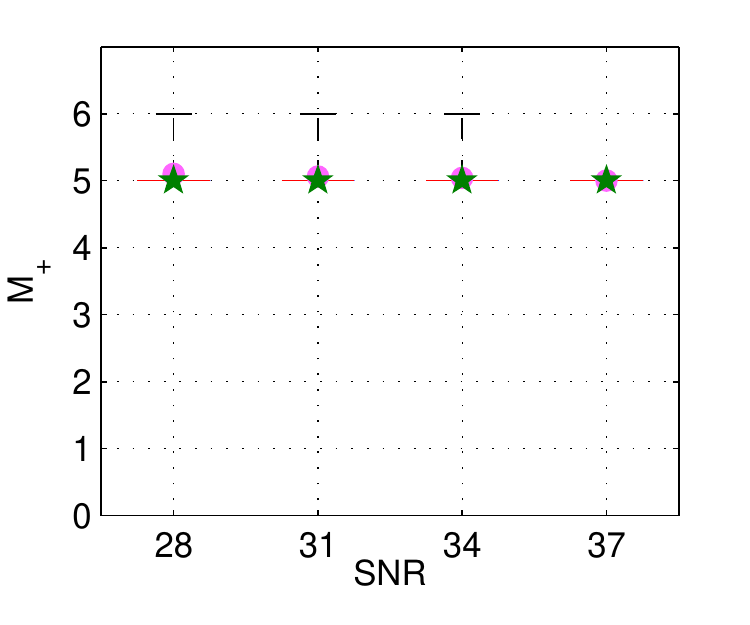}}
\subfloat[Recovered transmitters.]{\includegraphics[width=.25\textwidth]{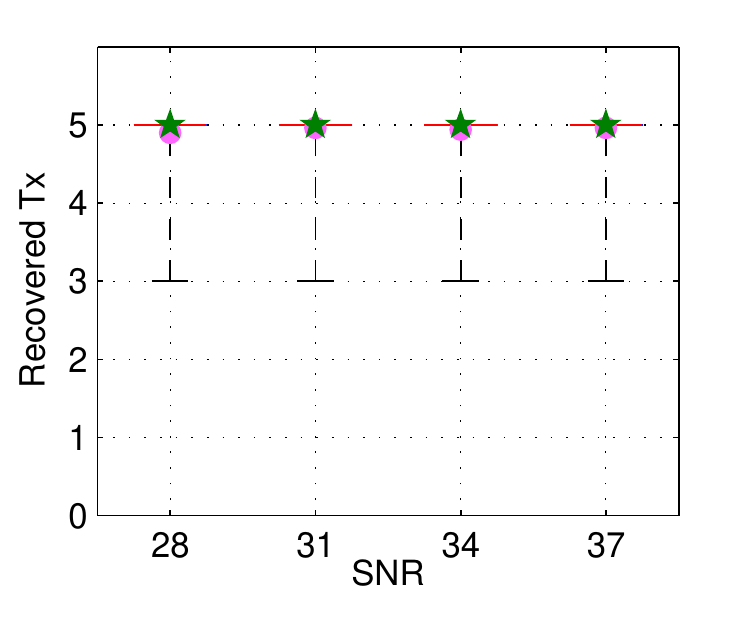}} \\ \vspace*{-10pt}
\subfloat[ADER.]{\includegraphics[width=.25\textwidth]{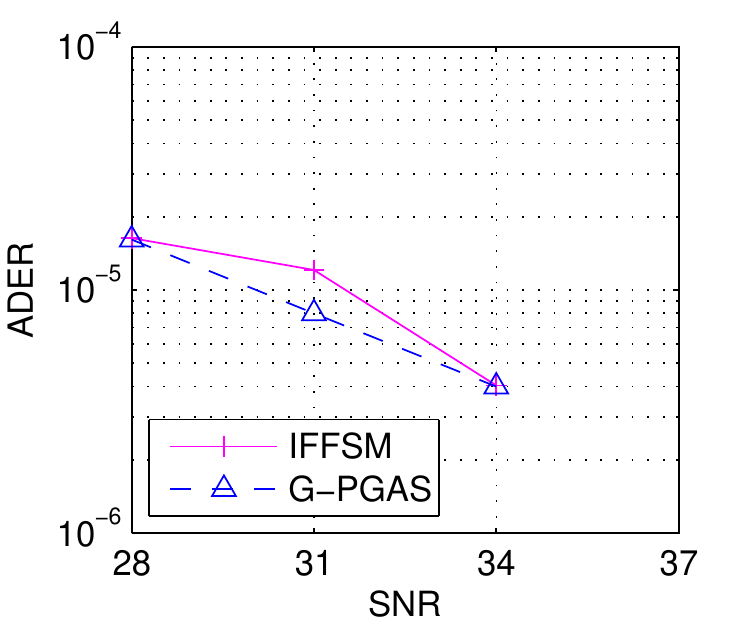}}
\subfloat[SER.]{\includegraphics[width=.25\textwidth]{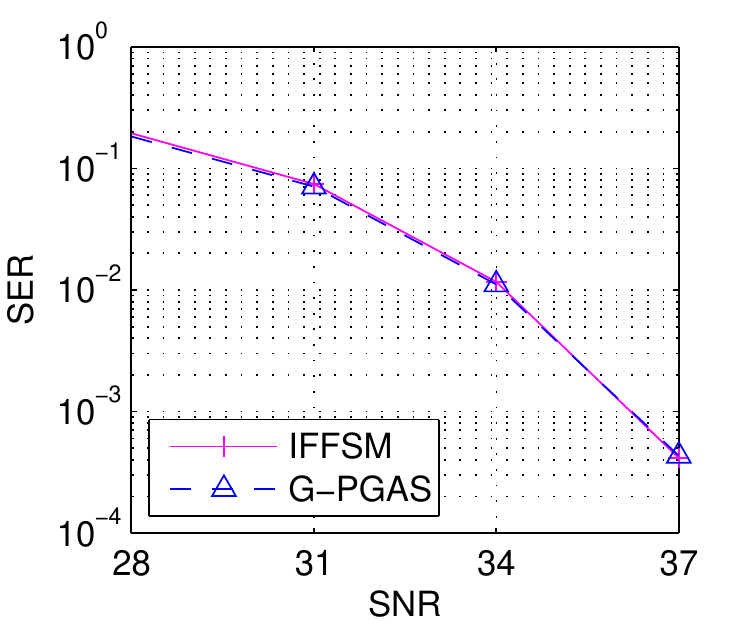}} \\ \vspace*{-10pt}
\subfloat[MSE.]{\includegraphics[width=.25\textwidth]{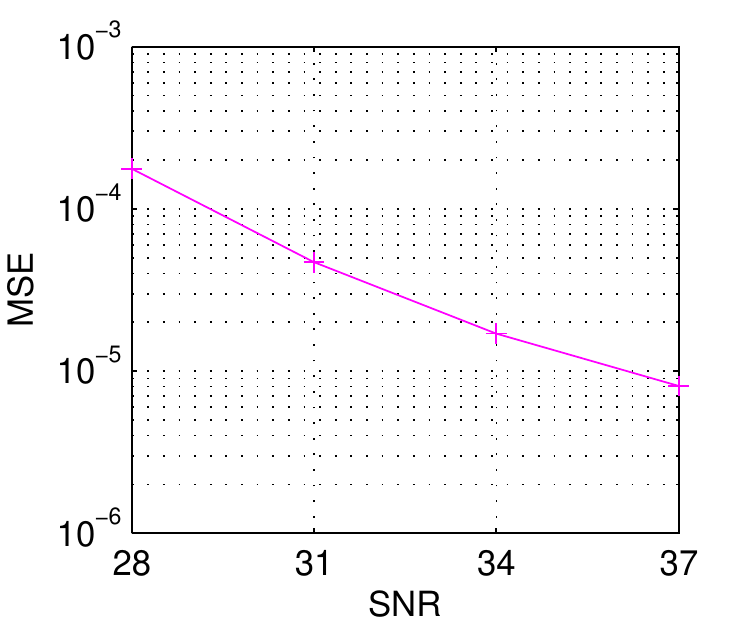}}
\caption{{Results for different SNRs using a 1024-QAM constellation.}}\label{ch6:fig:results_L1_M10} \vspace*{-10pt}
\end{figure}

In Figure~\ref{ch6:fig:results_Ltrue}, we show the obtained results for different values of the parameter $L$, ranging from $1$ to $5$. We use the true value of the channel length $L$ for inference, and we consider $\sigma_y^2=10^{0.9}$ in these experiments. The figure shows the {ADER}, the {SER}, the {MSE}, and box-plot representations of the inferred number of transmitters $M_+$ and the number of recovered transmitters. Here, it becomes clear that our model can exploit the redundancy introduced by the channel memory, as the performance in terms of SER and ADER improves as $L$ increases. The MSE also improves with $L$, although it reaches a constant value for $L>3$, similarly to the experiments in which we increase the number of receivers (although differently, in both cases we add redundancy to the observations). We can also observe that the performance is similar to the genie-aided methods (we do not run the FFBS algorithm for $L=5$ due to its computational complexity).

\begin{figure}[t]
\centering
\subfloat[Inferred $M_+$.]{\includegraphics[width=.25\textwidth]{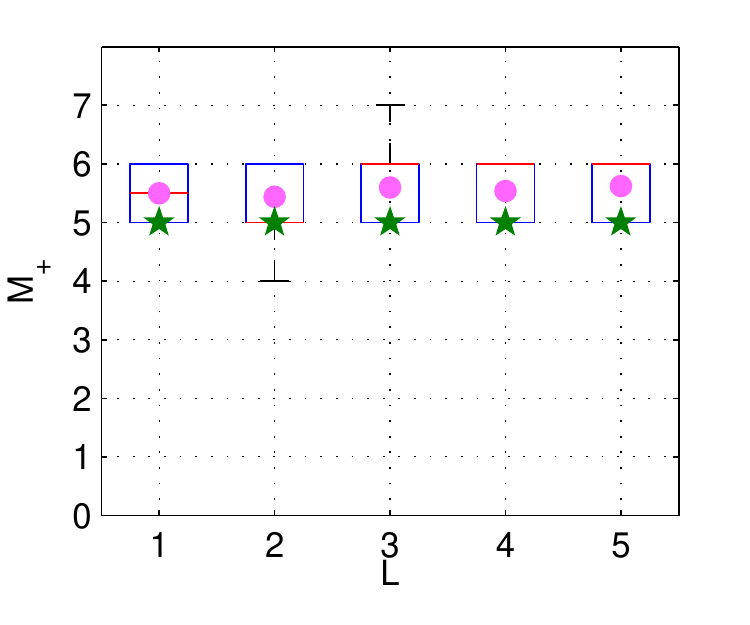}}
\subfloat[Recovered transmitters.]{\includegraphics[width=.25\textwidth]{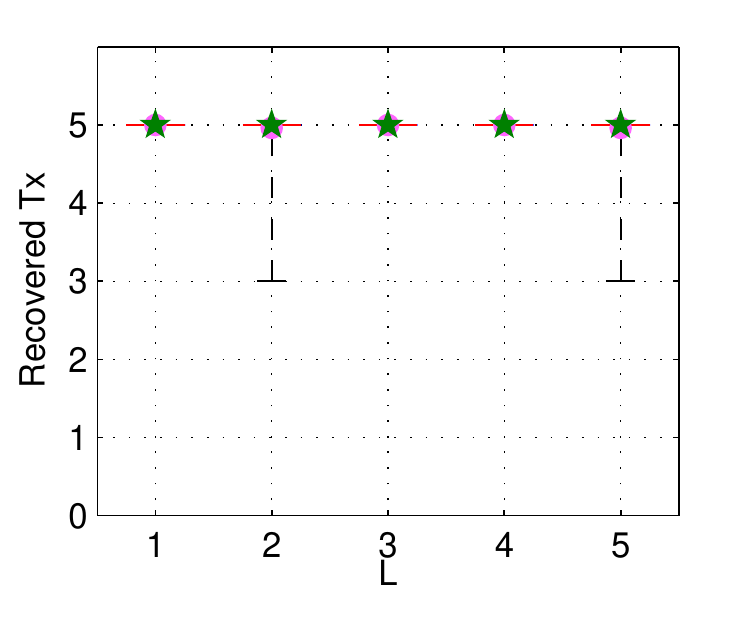}} \\ \vspace*{-10pt}
\subfloat[ADER.]{\includegraphics[width=.25\textwidth]{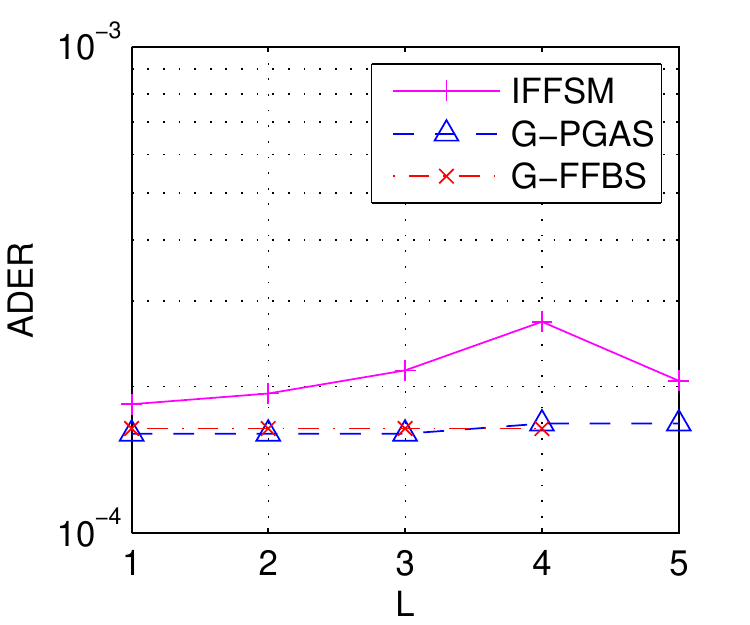}}
\subfloat[SER.]{\includegraphics[width=.25\textwidth]{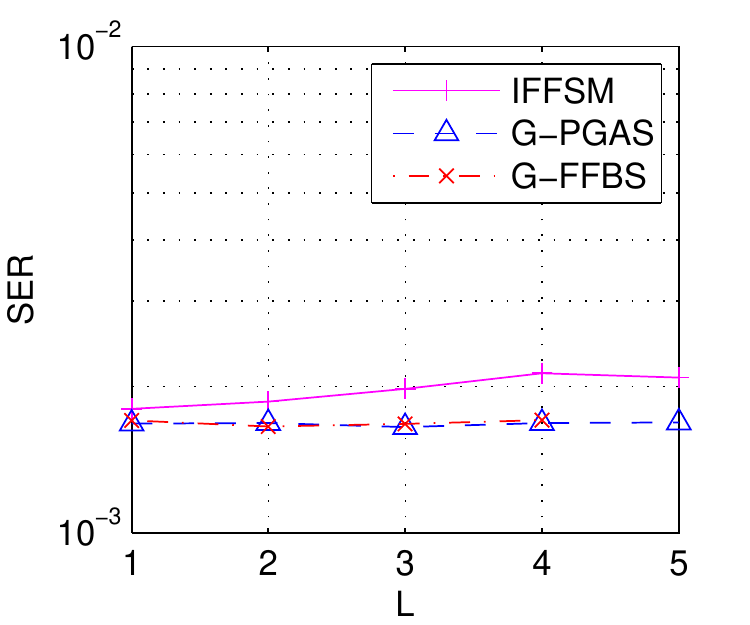}} \\ \vspace*{-10pt}
\subfloat[MSE.]{\includegraphics[width=.25\textwidth]{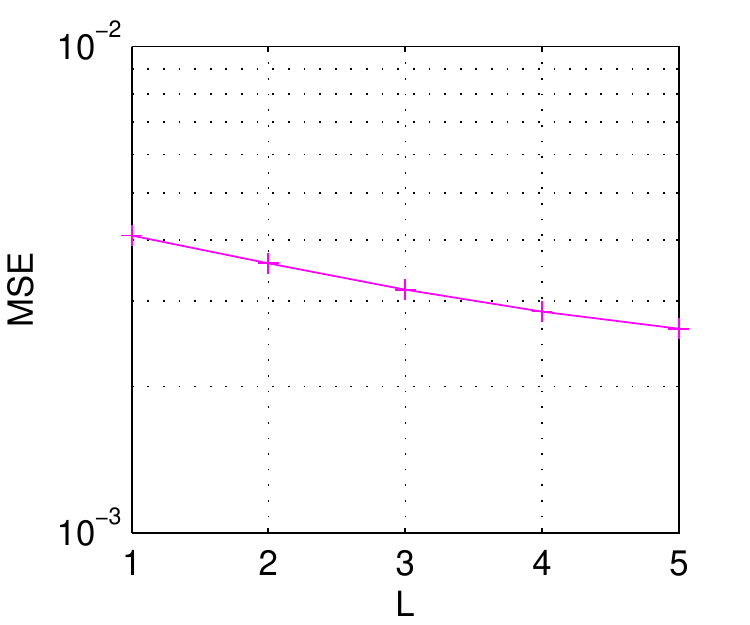}}
\caption{Results for different channel lengths ($L_\textrm{true}=1$).}\label{ch6:fig:results_sweepL} \vspace*{-10pt}
\end{figure}

Finally, we run an experiment with a different constellation order. Specifically, we use a 1024-QAM constellation, normalized to yield unit energy, and we vary the {noise variance} from $\s^2_y=10^{1.5}$ to $\s^2_y=10^{2.4}$ {(SNR from 28dB to 37dB)}. In order to improve the mixing of the sampling procedure, and due to the high cardinality of $\Acal$, after running $5000$ iterations of the inference algorithm, we run $20000$ additional iterations in which we sequentially sample each hidden chain conditioned on the current value of the remaining ones, similarly to the standard FFBS algorithm for factorial models. We proceed similarly for the genie-aided PGAS method. We do not run the genie-aided FFBS or BCJR methods due to their high computational complexity in this scenario. We plot in Figure~\ref{ch6:fig:results_L1_M10} the {ADER}, {SER}, and {MSE}, as well as a box-plot representation of the inferred number of transmitters $M_+$ and the number of recovered transmitters. We observe that, for the considered values of the SNR, we can recover the five transmitters in nearly all the cases. Furthermore, the SER curve of our approach closely follows the curve of the genie-aided method with perfect knowledge of the channel coefficients and the number of transmitters.

\xhdr{Results for mismatched channel length}
In the experiments above, we have assumed that the true value of the channel length is known at the receiver side. As this scenario might not be realistic, we now run an experiment to show that we can properly estimate the transmitted symbols and the channel coefficients as long as our inference algorithm considers a sufficiently large value of $L$. For this purpose, we use our base experimental setup and generate data using $L=1$ (i.e., memoryless channel). However, we use different values for the channel length $L$ for inference.

In Figure~\ref{ch6:fig:results_sweepL}, we show the obtained results for $L$ ranging from $1$ to $5$. The obtained {ADER} and {SER} do not significantly degrade with increasing values of $L$, and we are able to recover the five transmitters in nearly all the cases. Interestingly, the {MSE} improves as $L$ increases. This is a consequence of the way we measure it when $L$ is larger than the ground truth, as we compare our channel estimates with zero. The fact that the {MSE} becomes lower indicates that we obtain better estimates for the zero coefficients than for the non-zero ones, which in turn implies that our inference algorithm can properly decrease the channel variances $\sigma^2_\ell$ when needed.

\begin{figure}[t]
\centering
\includegraphics[width=.35\textwidth]{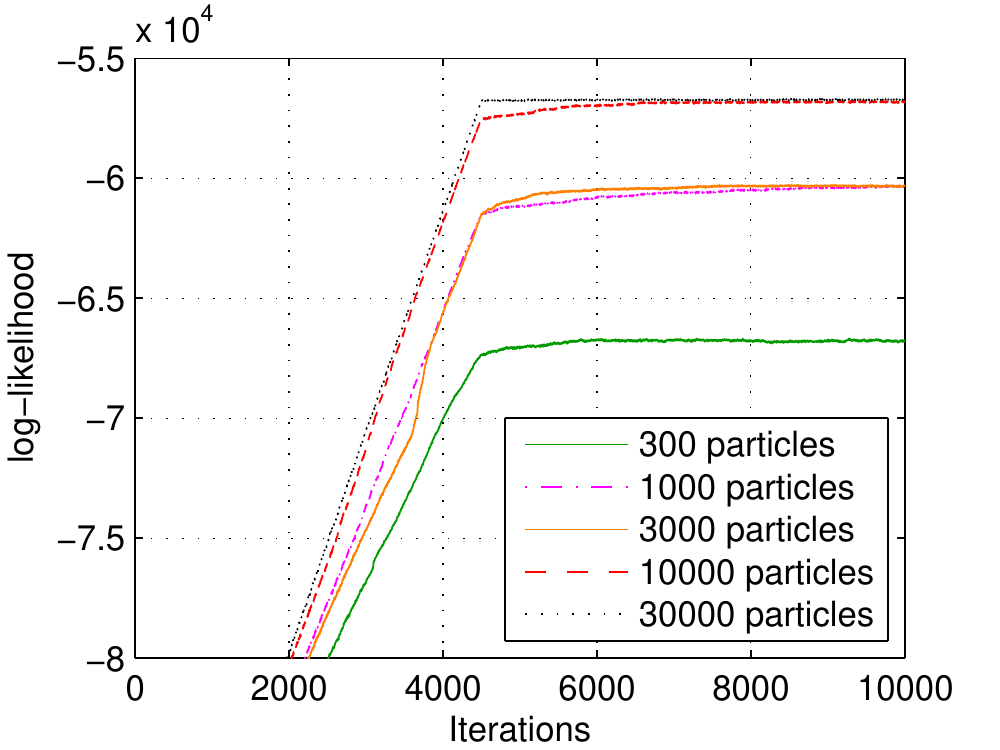}
\caption{Log-likelihood for varying number of particles ($L=1$). The initial slope is due to the tempering procedure, in which we linearly increase the SNR at each iteration.}\label{ch6:fig:resultsNpart1}
\end{figure}

\xhdr{Sensitivity to the number of particles}
Finally, we evaluate the impact of the number of particles in the PGAS kernel on the performance of the proposed algorithm. 
Note that, as the effective dimensionality of the hidden space increases, we should expect a larger number of particles to be required in order to properly estimate the transmitted symbols. To see this, we design an experiment with $10$ transmitters and $\sigma_y^2=10^{0.3}$. Figure~\ref{ch6:fig:resultsNpart1} shows the log-likelihood trace plot for $10000$ iterations of the inference algorithm, with a number of particles ranging from $300$ to $30000$. This experiment is based on a single run of the algorithm, as it is enough to understand its behavior. It can be seen that the best performance is achieved with the largest number of considered particles. Additionally, this plot suggests that $P=10000$ particles are enough for this scenario.

\begin{figure}[t]
\centering
\subfloat[Inferred $M_+$.]{\includegraphics[width=.26\textwidth]{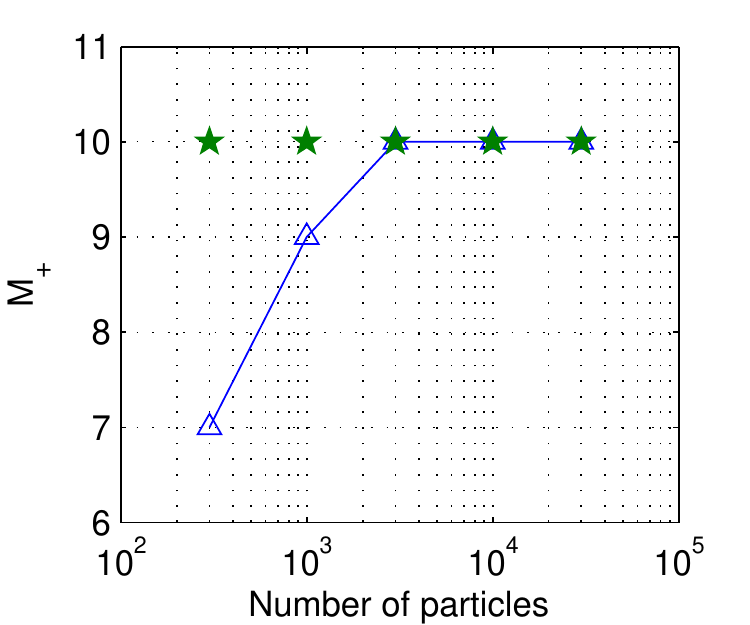}}\\
\subfloat[Recovered transmitters.]{\includegraphics[width=.5\textwidth]{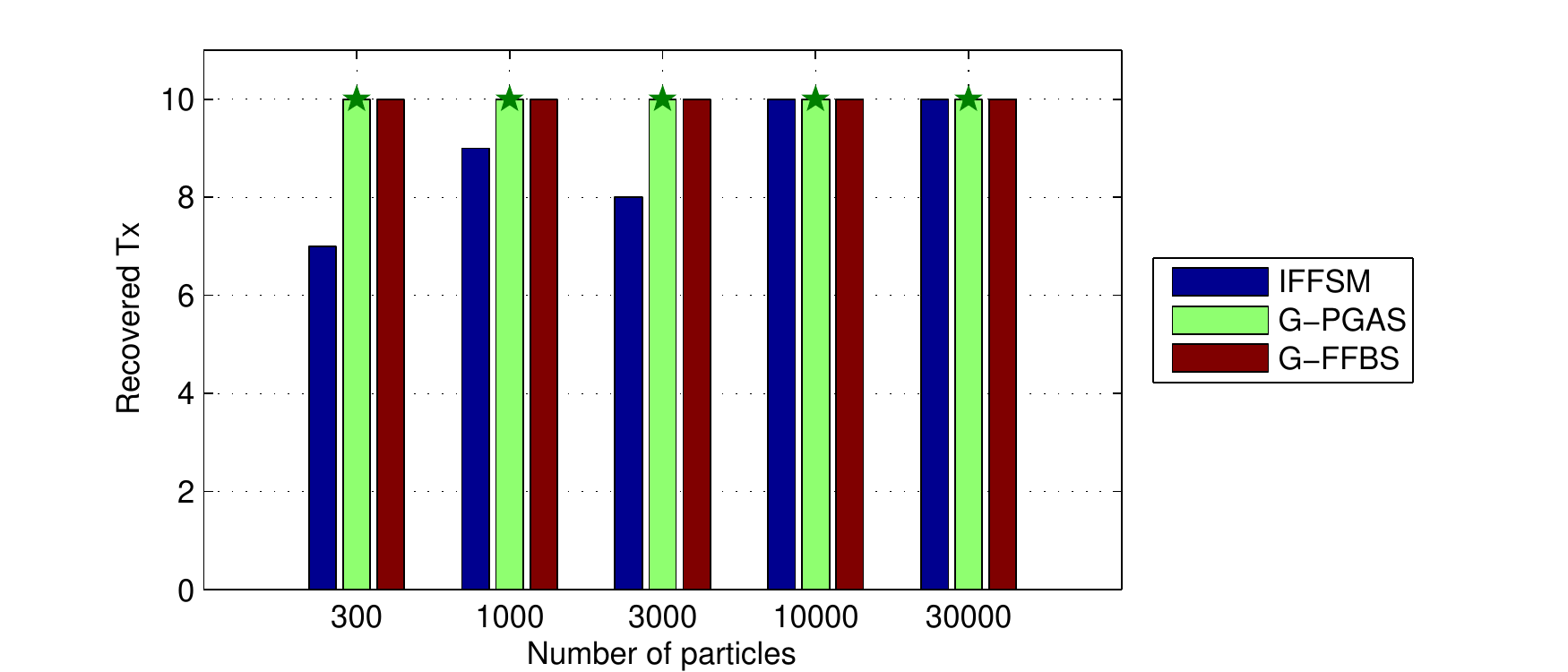}\label{ch6:fig:resultsNpart2b}}
\caption{Number of inferred and recovered transmitters for varying number of particles ($L=1$). The green star indicates the ground truth.}\label{ch6:fig:resultsNpart2}
\end{figure}

We also show in Figure~\ref{ch6:fig:resultsNpart2} the number of inferred transmitters $M_+$, as well as the number of recovered transmitters, for each value of $P$.  In this figure, we represent with a green star the true number of transmitters (again, these results are obtained after a single run of the algorithm). Although we infer $M_+=10$ transmitters with only $P=3000$ particles, Figure~\ref{ch6:fig:resultsNpart2b} shows that we only recover $8$ of them. The other two transmitters have been mixed in the last two inferred chains. In agreement with Figure~\ref{ch6:fig:resultsNpart1}, increasing the number of particles from $P=10000$ to $30000$ does not seem to improve performance: in both cases our algorithm is able to recover all the transmitters. Furthermore, even the genie-aided {PGAS} algorithm, which has perfect knowledge of the channel coefficients, needs a large value of $P$ (above $3000$) in order to recover all the transmitters.

We can conclude from these plots that we should adjust the number of particles based on the number of transmitters. However, the number of transmitters is an unknown quantity that we need to infer. There are two ways to overcome this apparent limitation. A sensible solution is to adaptively select the number of particles $P$ as a function of the current number of active transmitters, $M_+$. In other words, as we gather evidence for the presence of more transmitters, we consequently increase $P$. A second approach, which is computationally less demanding but may present poorer mixing properties, consists in running the {PGAS} inference algorithm sequentially over each chain, conditioned on the current value of the remaining transmitters, similarly to the standard {FFBS} procedure for the {IFHMM} \cite{VanGael2009}. Alternatively, we can apply the PGAS algorithm over fixed-sized blocks of randomly chosen transmitters.

\subsection{Ray-tracing Model}\label{subsec:realData}

With the aim of considering a more realistic communication scenario, we use WISE software \cite{Fortune1995} to design an indoor wireless system. This software tool, developed at Bell Laboratories, includes a 3D ray-tracing propagation model, as well as algorithms for computational geometry and optimization, to calculate measures of radio-signal performance in user-specified regions. Its predictions have been validated with physical measurements. 

\begin{figure}[b]
\centering
\includegraphics[width=.5\textwidth,clip=true,trim=3.2cm 12.8cm 1.7cm 12.7cm]{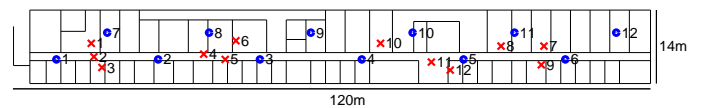}
\caption{Plane of the considered office building. Circles represent receivers, and crosses represent transmitters. All transmitters and receivers are placed at a height of $2$ metres.}\label{ch6:fig:planeOffice}
\end{figure}

Using WISE software and the map of an office located at Bell Labs Crawford Hill, we place $12$ receivers and $12$ transmitters across the office, intentionally placing the transmitters together in order to ensure that interferences occur in the nearby receivers. Figure~\ref{ch6:fig:planeOffice} shows the considered map.

We consider a Wi-Fi transmission system with a bandwidth of $20$ MHz or, equivalently, $50$ ns per channel tap. We simulate the transmission of $1000$-symbol bursts over this communication system, using a {QPSK} constellation normalized to yield unit energy. We scale the channel coefficients by a factor of $100$, and we consequently scale the noise variance by $10^4$, yielding $\sigma_y^2\approx 7.96\times 10^{-9}$. We set the transmission power to $0$ dBm. Each transmitter becomes active at a random point, uniformly sampled in the interval $[1,T/2]$. We set $T=2000$ time instants, so the burst duration of $1000$ symbols ensures overlapping among all the transmitted signals.

Wi-Fi systems are not limited by the noise level, which is typically small enough, but by the users' interferences, which can be avoided by using a particular frequency channel for each user. Our goal is to show that cooperation of receivers in a Wi-Fi communication system can help recover the symbols transmitted by several users even when they simultaneously transmit over the same frequency channel, therefore allowing for a larger number of users in the system.

In our experiments, we vary $L$ from $1$ to $5$. Five channel taps correspond to the radio signal travelling a distance of 750 m, which should be enough given the dimensions of this office space (the signal suffers attenuation when it reflects off the walls, so we should expect it to be negligible in comparison to the line-of-sight ray after a 750-m travelling distance). Following the tempering procedure described above, we initialize the algorithm with $\sigma_y^2\approx 15.85$ and we linearly increase the {SNR} for around $26600$ iterations, running $3400$ additional iterations afterwards. We compare our {IFFSM} with a non-binary {IFHMM} model with state space cardinality $|\Acal\bigcup \{0\}|^L$ using {FFBS} sweeps for inference (we do not run the {FFBS} algorithm for $L=5$ due to its computational complexity). In this case, we set the hyperparameters as $\sigma_H^2=0.01$, $\lambda=0.5$, $\kappa=1$, $\alpha=1$, $\beta_0=2$ and $\beta_1=0.1$.


We show in Table~\ref{ch6:tab:M_ESTcomm} the number of transmitters recovered after running the two inference algorithms, together with the inferred value of $M_+$, averaged for the last $2000$ iterations. We see that the {IFFSM} recovers the $12$ transmitters in all cases, and it does not tend to create spurious chains. In contrast, the {IFHMM} significantly overestimates the number of transmitters, which deteriorates the overall symbol estimates and, as a consequence, not all the transmitted symbols are recovered. In the best case, the {IFHMM} only recovers $6$ out of the $12$ transmitters. In the extreme case of $L=4$, the inference algorithm for the {IFHMM} estimates $57$ chains, but none of them corresponds to a true transmitter; rather, each transmitter is explained by a combination of several chains.

For completeness, we additionally report in Table~\ref{ch6:tab:MSE1comm} the {MSE} of the channel coefficients, averaged for the last $2000$ iterations. We sort the transmitters so that the {MSE} is minimized, and ignore the extra inferred transmitters
. As expected, the {MSE} decreases as we consider a larger value of $L$, since the model better fits the true radio propagation model.

\begin{table}[b]
\centering
\subfloat[\# Recovered transmitters / Inferred $M_+$.]{
\begin{tabular}{|c||c|c|c|c|c|}
\hline 
\multirow{2}{*}{Model} & \multicolumn{5}{ c| }{$L$} \\ \cline{2-6}
 & 1 & 2 & 3 & 4 & 5\\ \hline \hline
IFFSM & 12/13  & 12/12  & 12/12   & 12/12 & 12/12 \\ \hline
IFHMM & 6/31 & 6/20  &  6/21  & 0/57 & $-$  \\ \hline
\end{tabular}\label{ch6:tab:M_ESTcomm}
}
\quad
\subfloat[MSE of the channel coefficients ($\times 10^{-5}$).]{
\begin{tabular}{|c||c|c|c|c|c|}
\hline 
\multirow{2}{*}{Model} & \multicolumn{5}{ c| }{$L$} \\ \cline{2-6}
 & 1   & 2 & 3 & 4 & 5\\ \hline \hline
IFFSM & $8.13$ &  $8.09$ &  $1.39$  & $0.37$ & $0.25$ \\ \hline
IFHMM & $8.20$ &  $4.11$ &  $0.84$  & $-$ & $-$  \\ \hline
\end{tabular}\label{ch6:tab:MSE1comm}
}
\\
\caption{Results for the Wi-Fi experiment.}
\end{table}

%


\section{Conclusions and Discussion}\label{sec:conclusions}

In this paper, we have proposed a fully blind approach for joint channel estimation and detection of the transmitted data when the number of transmitters is unknown. Our approach is based on a BNP model, which we refer to as IFFSM, and considers a potentially unbounded number of stochastic finite-memory FSMs that evolve independently over time. Our model and the PGAS-based inference algorithm allow us to readily account for frequency-selective channels, avoiding the exponential complexity with respect to the channel length of previous approaches. We have evaluated the performance of our approach through a comprehensive experimental design using both synthetic and real data.

These results are promising for the suitability of BNPs applied to signal processing for communications. As a result, this paper opens several further research lines on this area. Some examples of such possible future research lines are:
\begin{itemize}
\item Improving the scalability of the inference algorithm. This would allow us to account for both a larger number of transmitters and larger observation sequences at a sub-linear computational cost. We believe that further investigations on faster approximate inference algorithms would be of great interest and importance for the development of practical receivers that operate in real-time.
\item An online inference algorithm. In many cases, the receiver does not have access to a fixed window of observations, but data arrives instead as a never-ending stream. Thus, an online inference algorithm instead of a batch one is also of potential interest.
\item Modelling of time-varying channels. Our current approach is restricted to static channels, i.e., the channel coefficients do not vary over time. A potentially useful research line may consist in taking into account the temporal evolution of the channel coefficients.
\item An extension of the model that accounts for coding schemes. A channel coding scheme can be used in order to add redundancy to the user's data, effectively decreasing the resulting bit error probability. This redundancy can potentially be included into the model and exploited by the inference algorithm.
\item {An extension for sparse code multiple access (SCMA). Our approach can also be adapted to the particularities of the uplink SCMA techniques} \cite{Au2014,Bayesteh2014}{, which use non-orthogonal transmission signals in the random access channel to reduce latency. The signal structure and the sparsity of the code/pilot scheme may simplify the inference procedure of the Bayesian model, at the cost of an upper bound on the number of potential users in the network. This would result in a model that does not need to be nonparametric, but our joint approach may still reduce the errors of the current several-stage detectors.}
\end{itemize}

An extension of our IFFSM, less related to the aforementioned research lines, consists in a semi-Markov approach. In this way, transmitters are assumed to send only a burst of symbols during the observation period, and a changepoint detection algorithm may detect the (de)activation instants.

\section*{Acknowledgments}
Francisco J.\ R.\ Ruiz acknowledges the European Union H2020 program (Marie Sk\l{}odowska-Curie grant agreement 706760).



\bibliographystyle{IEEEtran}
\bibliography{Bib,moreBib}


%





\ifCLASSOPTIONcaptionsoff
  \newpage
\fi



%



%

\begin{IEEEbiography}[{\includegraphics[width=1in,height=1.25in,clip,keepaspectratio]{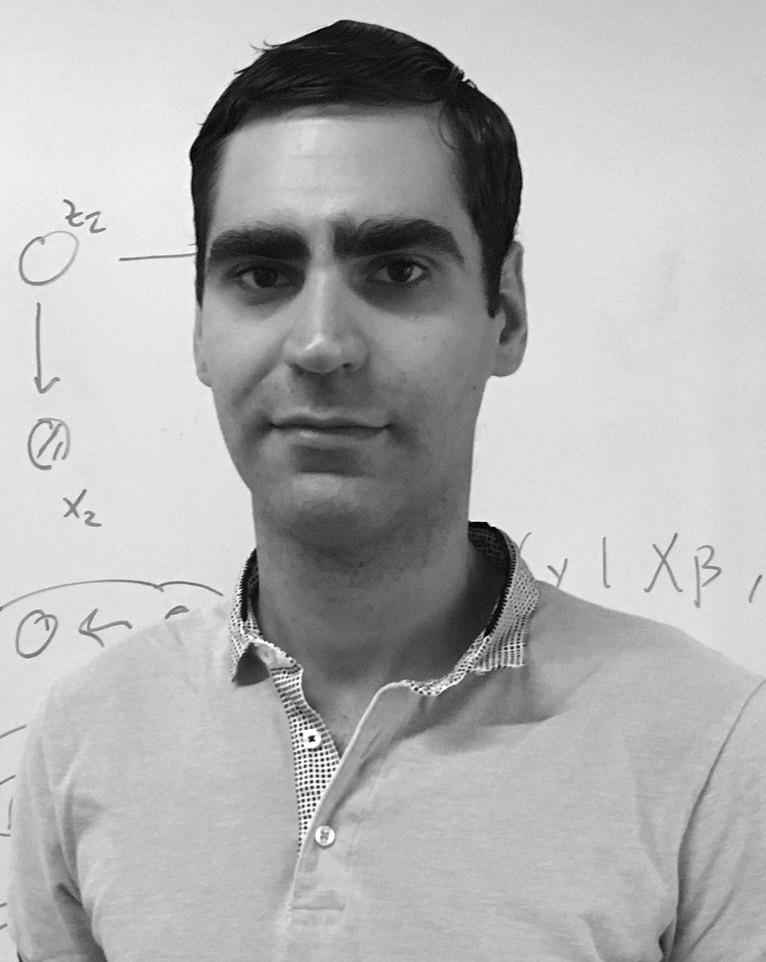}}]{Francisco J.\ R.\ Ruiz}
received the B.Sc.\ degree in Electrical Engineering from the University of Seville, Spain, and the M.Sc.\ degree in Multimedia and Communications from the University Carlos III in Madrid, Spain, in 2010 and 2012, respectively. He received the Ph.D.\ degree in 2015 from the University Carlos III in Madrid. He is currently a Postdoctoral Research Scientist at the Department of Computer Science in Columbia University, and at the Engineering Department in the University of Cambridge. Francisco holds a Marie Sk\l{}odowska-Curie fellowship in the context of the E.U.\ Horizon 2020 program. His research is focused on statistical machine learning, specially Bayesian modeling and inference.
\end{IEEEbiography}

\begin{IEEEbiography}[{\includegraphics[width=1in,height=1.25in,clip,keepaspectratio]{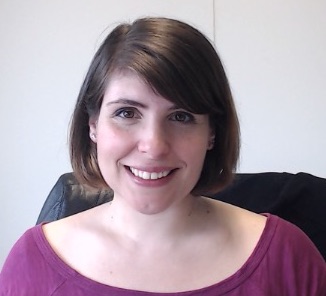}}]{Isabel Valera}
is currently leading the Probabilistic Learning Group in the Department of Empirical Inference at the Max Planck Institute for Intelligent Systems in T\"{u}bingen. Prior to this she was a post-doctoral researcher at the University of Cambridge, and previously, she was a Humboldt post-doctoral fellowship holder at Max Planck Institute for Software Systems. She obtained her PhD in 2014 and her Master degree in 2012, both from the University Carlos III in Madrid.
\end{IEEEbiography}

\begin{IEEEbiography}[{\includegraphics[width=1in,height=1.25in,clip,keepaspectratio]{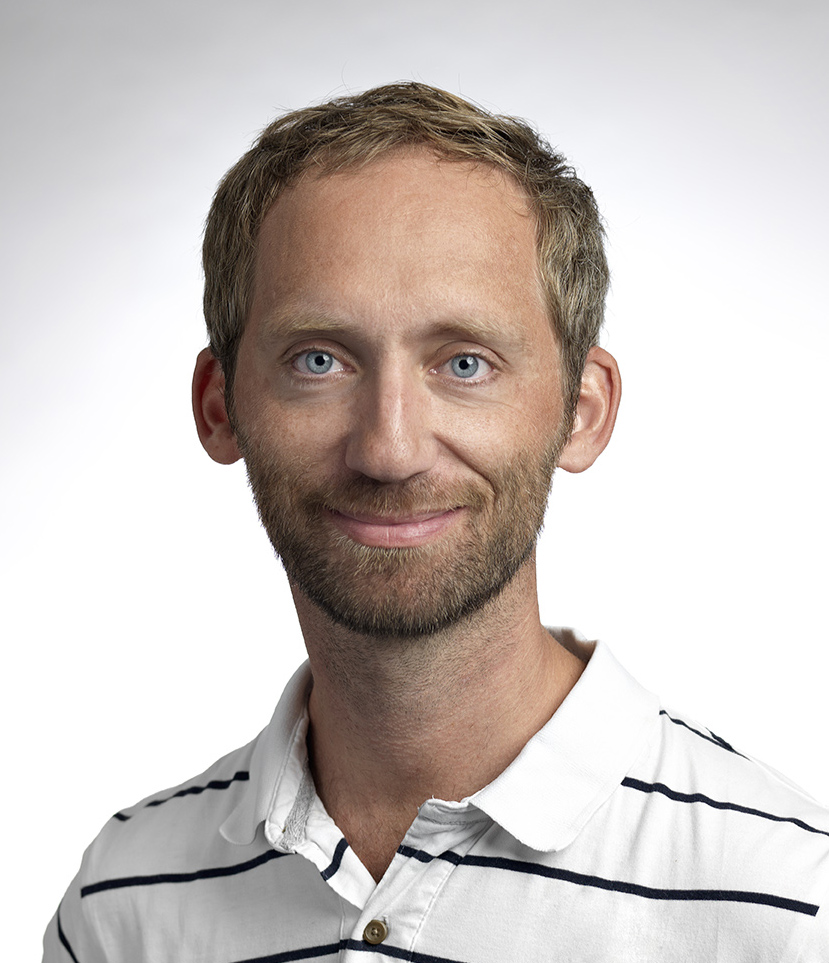}}]{Lennart Svensson}
was born in \"{A}lv\"{a}ngen, Sweden in 1976. He received the M.S.\ degree in electrical engineering in 1999 and the Ph.D.\ degree in 2004, both from Chalmers University of Technology, Gothenburg, Sweden. He is currently Professor of Signal Processing, again at Chalmers University of Technology. His main research interests include machine learning and Bayesian inference in general, and nonlinear filtering, deep learning, and multi-target tracking in particular.
\end{IEEEbiography}

\vfill

\newpage

\begin{IEEEbiography}[{\includegraphics[width=1in,height=1.25in,clip,keepaspectratio]{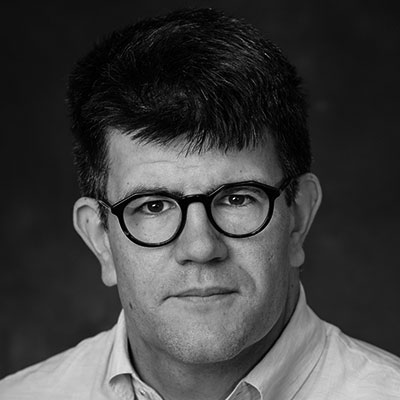}}]{Fernando Perez-Cruz}
(IEEE Senior Member) was born in Sevilla, Spain, in 1973. He received a PhD.\ in Electrical Engineering in 2000 from the Technical University of Madrid and an MSc/BSc in Electrical Engineering from the University of Sevilla in 1996. He is the Chief Data Scientist at the Swiss Data Science Center (ETH Zurich and EPFL). He has been a member of the technical staff at Bell Labs and an Associate Professor with the Department of Signal Theory and Communication at University Carlos III in Madrid and Computer Science at Stevens Institute of Technology. He has been a visiting professor at Princeton University under a Marie Curie Fellowship and a Research Scientist at Amazon. He has also held positions at the Gatsby Unit (London), Max Planck Institute for Biological Cybernetics (T\"{u}bingen), BioWulf Technologies (New York) and the Technical University of Madrid and Alcala University (Madrid). His current research interest lies in machine learning and information theory and its application to signal processing and communications. Fernando has organized several machine learning, signal processing, and information theory conferences. Fernando has supervised 8 PhD students and numerous MSc students, as well as one junior and one senior Marie Curie Fellow.
\end{IEEEbiography}

\vfill









\end{document}